\documentclass[twocolumn,iop,apj,twocolappendix]{aastex63}
\usepackage{amsfonts,amsmath,graphicx,natbib,url,hyperref,nicefrac}
\usepackage{amssymb, verbatim, subfigure, paralist, soul, comment}
\usepackage{float, tabularx, booktabs}
                    
\usepackage{verbatim}
\usepackage{tikz}
\usepackage{color}

\usepackage[T1]{fontenc}
\usepackage{aecompl}
\usepackage{calligra}
\DeclareMathAlphabet{\mathcalligra}{T1}{calligra}{m}{n}
\DeclareFontShape{T1}{calligra}{m}{n}{<->s*[2.2]callig15}{}

\usepackage{hyperref}
\hypersetup{
    pdfnewwindow=true,      % links in new window
    colorlinks=true,       % false: boxed links; true: colored links
    linkcolor=orange,          % color of internal links
    citecolor=cyan,        % color of links to bibliography
    filecolor=blue,      % color of file links
    urlcolor=blue           % color of external links
}

\def\apjl{ApJL}               % Astrophysical Journal, Letters
\def\apjs{ApJS}               % Astrophysical Journal, Supplement
\def\aap{A\&A}                % Astronomy and Astrophysics
\DeclareMathAlphabet{\mathcalligra}{T1}{calligra}{m}{n}
\DeclareFontShape{T1}{calligra}{m}{n}{<->s*[2.2]callig15}{}

\newcommand{\scripty}[1]{\ensuremath{\mathcalligra{#1}}}

\def\scr{\scripty{r}}
\def\eg{e.g.}

\usepackage{color}

\usepackage{comment}

\shorttitle{Santa Barbara Binary} 
\shortauthors{Duffell et al.}

\begin{document}

\title[]{The Santa Barbara Binary-Disk Code Comparison}

\author[0000-0001-7626-9629]{Paul C. Duffell}
\affiliation{Department of Physics and Astronomy, Purdue University, 525 Northwestern Avenue, West Lafayette, IN 47907, USA}
\email{pduffell@purdue.edu}

\author[0000-0001-6157-6722]{Alexander J. Dittmann}
\affiliation{Department of Astronomy and Joint Space-Science Institute, University of Maryland, College Park, MD 20742-2421, USA}
\affiliation{Theoretical Division, Los Alamos National Laboratory, Los Alamos, NM 87545, USA}
\email{dittmann@umd.edu}

\author[0000-0002-1271-6247]{Daniel J. D'Orazio}
\affiliation{Niels Bohr International Academy, Niels Bohr Institute, Blegdamsvej 17, 2100 Copenhagen, Denmark}
\email{daniel.dorazio@nbi.ku.dk}

\author[0000-0002-8400-0969]{Alessia Franchini}
\affiliation{Universität Zürich, Institut für Astrophysik, Winterthurerstrasse 190, CH-8057 Zürich, Switzerland}
\affiliation{Dipartimento di Fisica "G. Occhialini", Universit\`a degli Studi di Milano-Bicocca, Piazza della Scienza 3, I-20126 Milan, Italy}
\affiliation{INFN, Sezione di Milano-Bicocca, Piazza della Scienza 3, I-20126 Milano, Italy}

\author[0000-0001-5253-1338]{Kaitlin M. Kratter}
\affiliation{University of Arizona, 933 N Cherry Ave, Tucson, AZ 85721, USA}

\author[0000-0002-8873-6826]{Anna B.T. Penzlin}
\affiliation{Institut f\"ur Astronomie und Astrophysik, Universität T\"ubingen,
Auf der Morgenstelle 10, D-72076 T\"ubingen, Germany}
\affiliation{Astrophysics Group, Department of Physics, Imperial College London, Prince Consort Rd, London, SW7 2AZ, UK}

\author[0000-0001-5378-7749]{Enrico Ragusa}
\affiliation{Dipartimento di Matematica, Università degli Studi di Milano, Via Saldini 50, 20133, Milano, Italy}
\affiliation{ENS de Lyon, CRAL UMR5574, Universite Claude Bernard Lyon 1, CNRS, Lyon, F-69007, France}

\author[0000-0002-1530-9778]{Magdalena Siwek}
\affiliation{Center for Astrophysics, Harvard University, Cambridge, MA 02138, USA}

\author[0000-0002-3820-2404]{Christopher Tiede}
\affiliation{Niels Bohr International Academy, Niels Bohr Institute, Blegdamsvej 17, 2100 Copenhagen, Denmark}

\author[0000-0001-7167-6110]{Haiyang Wang}
\affiliation{Fudan University, Department of Physics, Shanghai 200433, China}
\affiliation{Department of Applied Mathematics and Theoretical Physics, University of Cambridge, Wilberforce Road, Cambridge CB3 0WA, UK}
\affiliation{TAPIR, California Institute of Technology, Pasadena, CA 91125, USA}

\author[0000-0002-1895-6516]{Jonathan Zrake}
\affiliation{Department of Physics and Astronomy, Clemson University, Clemson, SC 29634, USA}

\author[0000-0001-8291-2625]{Adam M. Dempsey}
\affiliation{X Computational Physics Division, Los Alamos National Laboratory, Los Alamos, NM 87545, USA}

\author[0000-0003-3633-5403]{Zoltan Haiman}
\affiliation{Columbia University, Departments of Astronomy and Physics, 550 West 120th Street, New York, NY, 10027, USA}

\author[0000-0001-6106-7821]{Alessandro Lupi}
\affiliation{Dipartimento di Scienza e Alta Tecnologia, Universit\`a degli Studi dell'Insubria, via Valleggio 11, I-22100 Como, Italy}
\affiliation{Dipartimento di Fisica "G. Occhialini", Universit\`a degli Studi di Milano-Bicocca, Piazza della Scienza 3, I-20126 Milan, Italy}
\affiliation{INFN, Sezione di Milano-Bicocca, Piazza della Scienza 3, I-20126 Milano, Italy}

\author[0000-0002-3164-5923]{Michal Pirog}
\affiliation{Wilkes Honors College, Florida Atlantic University, Jupiter, FL 33458, USA}
\affiliation{Department of Physics and Astronomy, West Virginia University, Morgantown, WV 26506, USA}
\affiliation{Center for Gravitational Waves and Cosmology, West Virginia University, Chestnut Ridge Research Building, Morgantown, WV 26505, USA}

\author[0000-0001-9068-7157]{Geoffrey Ryan}
\affiliation{Perimeter Institute for Theoretical Physics, 31 Caroline Street North, Waterloo, ON N2L 2Y5, Canada}

\begin{abstract}
We have performed numerical calculations of a binary interacting with a gas disk, using eleven different numerical methods and a standard binary-disk setup.  The goal of this study is to determine whether all codes agree on a numerically converged solution, and to determine the necessary resolution for convergence and the number of binary orbits that must be computed to reach an agreed-upon relaxed state of the binary-disk system.
We find that all codes can agree on a converged solution (depending on the diagnostic being measured).  The zone spacing required for most codes to reach a converged measurement of the torques applied to the binary by the disk is roughly 1\% of the binary separation in the vicinity of the binary components.
For our disk model to reach a relaxed state, codes must be run for at least 200 binary orbits, corresponding to about a viscous time for our parameters, $0.2 (a^2 \Omega_B /\nu)$ binary orbits, where $\nu$ is the kinematic viscosity.  
%We did not investigate dependence on binary mass ratio, eccentricity, disk temperature, or disk viscosity; therefore, these benchmarks may act as guides towards expanding converged solutions to the wider parameter space but might need to be updated in a future study that investigates dependence on system parameters.  
The largest discrepancies between codes resulted from the dimensionality of the setup (3D vs 2D disks).  We find good agreement in the total torque on the binary between codes, although the partition of this torque between the gravitational torque, orbital accretion torque, and spin accretion torque depends sensitively on the sink prescriptions employed.  In agreement with previous studies, we find a modest difference in torques and accretion variability between 2D and 3D disk models.  We find cavity precession rates to be appreciably faster in 3D than in 2D.
\end{abstract}

\section{Introduction}
Over the past few decades, the general astrophysical hydrodynamics problem of a disk interacting with a binary has been investigated in numerous studies.  It has been studied in the context of a stellar binary interacting with the protostellar disk \citep[\eg,][]{Bate1995, Dunhill+2015, Terquem+2015, MartinLubow:2019, Franchini2019}, 
and in the context of supermassive black hole binaries at sub-parsec separations interacting in galactic nuclei \citep[\eg,][]{ArmNat:2002:ApJL, Hayasaki:2007, MacFadyen:2008, Cuadra:2009, dittmann_decoupling, Krauth+2023}
It has been studied in 2D, in the plane of the binary orbit \citep[\eg,][]{Ochi+2005, Hanawa+2010, MacFadyen:2008, DHM:2013:MNRAS, YoungBairdCLarke:2015, Farris:2014, D'Orazio:CBDTrans:2016, MunozLai:2016, MirandaLai+2017, munoz19, duffell20, dittmann_sinks, dittmann_mach, penzlin_mach, siwek_ecc_orb,siwek_ecc, MaheshMcWillPirog:2023, TiedeDOrazio:2023, CimmermanRafikov:2023} and in 3D 
\citep[\eg,][]{AL94, ArtyLubow:1996, BateBonnell:1997, Hayasaki:2007,2011PhDT........36K, NixonRetro:2011,RagusaLodato:2016,ragusa2020,moody19, heath_nixon, Franchini2022, Franchini2023,2023arXiv231117144B,2023arXiv230714562P}, including magnetic fields \citep[\eg,][]{ ShiKrolik:2012, ShiKrolik:2015, BankertKrolikShoi:2015}, self-gravity \citep[\eg,][]{Cuadra:2009, Roedig:2011:eccevo, Roedig:2012:Trqs, Franchini2021,2023arXiv231117144B}, and general relativity \citep[e.g.,][]{FarrisGold:2012,Noble+2012,Gold:GRMHD_CBDII:2014,Bowen+2019,noble21,Combi+2022,Avara+2023,Mignon-Risse+2023}. It has been studied using a locally isothermal equation of state by many of these studies, and using a cooled disk with more involved thermodynamic assumptions in others \citep[\eg,][]{Farris:2015:Cool, d'Ascoli+2018, WesternacherSchneider2022, Combi+2022, Avara+2023, Krauth+2023}. 
What all of these studies have in common is that they employ numerical codes to integrate a set of hydrodynamic equations, to compute the gravitational interaction of a disk with a binary.

Many of these studies have had results which (at least seemingly) disagreed with one another; for example, it was often thought that disk-binary interaction would always remove angular momentum from the binary (essentially exerting a "drag force"), shrinking the binary separation \cite[e.g.,][]{DAngelowLubow:2008}.  However, with the advent of modern high-resolution hydrodynamical simulations, several studies have found net {\em positive} torque on the binary, enough to cause binary expansion. Circumbinary disk-driven outspirals were first observed in \citet{MirandaLai+2017}, and later confirmed by numerous other studies \citep[e.g.][]{munoz19,moody19, duffell20}. 
Subsequent studies have found that binary orbital evolution depends on the disk and binary parameters, such as the disk aspect ratio $h/r$ \citep{Tiede:2020,dittmann_mach,penzlin_mach,dittmann_q} the disk viscosity \citep{MirandaLai+2017,Franchini2022,dittmann_mach, penzlin_mach,dittmann_q}, the cooling timescale \citep{Sudarshan+2022, WangBaiLai:2022, wang23,2024arXiv240200938F}, and the binary eccentricity \citep{MirandaLai+2017,dorazio_ecc, siwek_ecc_orb}. \citet{LaiMunoz_rev:2023} recently presented a review summarizing the outcomes of many of these studies. 

This rapid influx of results is compounded by the fact that each study uses a different code, at different resolutions, integrating for different amounts of time, using different numerical choices such as boundaries, floors, and sinks, and including different amounts of additional physics (such as 3D and magnetic fields).  In order to help make sense of all of this, we present a code test for a uniform binary-disk setup, to determine just how closely our different codes agree, and what numerical ingredients (such as resolution and length of time integration) are necessary for convergence (or at least for agreement) between codes.

Nearly identical initial conditions are implemented across eleven different numerical schemes, running them at different resolutions and with different numerical choices.  We compute a few different measurements of the gravitational torque on the binary, the morphology of the cavity, and the accretion rate onto each binary component, in order to determine under what conditions the codes agree for different measured quantities.

The hydrodynamical setup is designed to be relatively simple while maintaining the essential ingredients of a time-dependent binary potential interacting with a disk.  In this way we will think of it as the "minimal" binary-disk problem.  The obvious advantage of using a minimal setup is that it can be tested across many different codes.  Additionally, performing these calculations in 2D enables codes to be run at higher resolution and for longer timescales than in 3D, which will make it possible to determine when the solution is converged.  Such a criterion might then be used to inform whether 3D studies are converged.  Another advantage of such a minimal problem is that it provides a straightforward way to test new codes on their ability to describe a binary-disk setup.  Our code outputs have been made public online at \url{physics.purdue.edu/duffell/data.html} and on Zenodo \cite{sbb_repository}.

\section{Codes}

A wide variety of codes will be compared on this binary setup.  It is worth noting that many of them are not being utilized up to their full potential; many of these codes are capable of 3D, MHD, self-gravity, general relativity, complex thermodynamics, and more.  Many of these capabilities need to be "turned off" in order to make an accurate comparison in a minimal binary-disk problem.   For each code, results from at least three numerical calculations at different resolutions will be presented and compared through the usage of a number of diagnostics described in Sec. \ref{sec:diagnostics}.

We list the essential characteristics of each code, as they are used in this study, in Table \ref{tab:hydro_codes}. Some codes, namely \texttt{Disco} and \texttt{Athena++}, have been used in multiple configurations.
For the sake of brevity, we defer comprehensive descriptions of each code to Appendix \ref{sec:codeAppendix}.

\begin{deluxetable*}{lccccc}
\tablenum{1}
\tablecaption{Overview of various hydrodynamics codes and their configurations used in this study. \label{tab:hydro_codes}}
\tablewidth{0pt}
\tablehead{
\colhead{Code} & \colhead{2D/3D} & \colhead{Numerical Scheme} & \colhead{\hspace{-0.5cm}Convergence Order} & \colhead{Excised Binary}
& \colhead{Discretization} }
\startdata
\texttt{Arepo}\tablenotemark{a}             & 2D             & Moving-mesh finite volume         & Second                     & No   & Voronoi tessellation          \\
\texttt{Athena++}\tablenotemark{b}          & 2D             & Finite volume                     & Third                      & No   & Cartesian          \\
\texttt{Athena++} (Excised)& 2D             & Finite volume                     & Second                     & Yes  & Cylindrical polar  \\
\texttt{Athena++} (3D)     & 3D             & Finite volume                     & Second                      & No   & Cartesian          \\
\texttt{Disco}\tablenotemark{c}             & 2D             & Moving-mesh finite volume         & Second                     & No   & Cylindrical polar  \\
\texttt{Disco} (Excised)   & 2D             & Moving-mesh finite volume         & Second                     & Yes  & Cylindrical polar  \\
\texttt{Fargo3D}\tablenotemark{d}            & 2D             & \hspace{-0.3cm}Orbital advection finite difference and volume   & Second                     & Yes  & Cylindrical polar  \\
\texttt{Gizmo}\tablenotemark{e}              & 3D             & Mesh-free finite mass             & Second                     & No   & Particles                \\
\texttt{Mara3}\tablenotemark{f}               & 2D             & Finite volume                     & Second                     & No   & Cartesian          \\
\texttt{Phantom}\tablenotemark{g}           & 3D             & Smoothed-particle hydrodynamics   & Second                        & No   & Particles                \\
\texttt{Pluto}  \tablenotemark{h}            & 2D             & Finite volume                     & Second                     & No   & Cylindrical polar  \\
\texttt{Sailfish}\tablenotemark{i}           & 2D             & Finite volume                     & Second                     & No   & Cartesian          \\
\enddata
\tablenotemark{a}{\citep{arepo,2013MNRAS.428..254M}}
\tablenotemark{b}{\citep{2020ApJS..249....4S}}
\tablenotemark{c}{\citep{DuffellMHDDISCO:2016}}
\tablenotemark{d}{\citep{FARGO3D}}
\tablenotemark{e}{\citep{Hopkins2015,Franchini2022}}
\tablenotemark{f}{\citep{Mara}}
\tablenotemark{g}{\citep{Price2018}}
\tablenotemark{h}{\citep{2007Mignone}}
\tablenotemark{i}{\citep{Westernacher-Schneider:2023}}
\caption{A summary of the different codes used in this study. We provide additional details for each in Appendix \ref{sec:codeAppendix}.
}
\end{deluxetable*}

\section{Test Problem}

\subsection{Binary Orbit}

We study an equal-mass binary with total binary mass $M$ (individual component masses $M_1 = M_2 = M/2$), on a fixed, circular ($e=0$) Keplerian orbit with semi-major axis $a$.  The binary angular frequency is denoted $\Omega_B$ while the disk angular frequency is $\Omega$.  For concreteness, the binary is initialized along the x-axis, orbiting in the x-y plane counter-clockwise when viewed from the positive z-axis.

%Define the disk aspect ratio as $h \equiv H/r \equiv 1/\mathcal{M}$ for disk scale height H, and radius $r$ measured from the binary barycenter.

Gravitational softening is implemented uniformly across all codes.  The binary potential can be expressed as
\begin{equation}
    \Phi_j = -\frac{G M_j}{\sqrt{ |\scr_{ij}|^2 + \epsilon^2 }},
\end{equation}
where $|\scr_{ij}|$ is the distance between $i^{th}$ gas cell/particle and $j^{th}$ binary component and $\epsilon$ is the gravitational softening length.
In the fiducial runs, we set $\epsilon=0.05a$, but we also explore its impact on the results in Section \ref{sec:eps}.  In that section, we demonstrate the binary torque to be insensitive to the gravitational softening for $\epsilon < 0.15 a$.

Most of the codes have employed "code units" where $GM = 1$, $a = 1$ (so that $\Omega_B = 1$), and $\Sigma_0 = 1$ ($\Sigma_0$ is defined in the initial conditions).  Codes that do not work in these units will normalize their results (e.g. reporting $\Sigma/\Sigma_0$ instead of $\Sigma$).  For example, simulations employing a "live binary" have an additional mass scale (the solution can depend on the normalization of $\Sigma_0$) and therefore such codes must set $\Sigma_0 \ll 1$ in order to find agreement with fixed-binary results in the "test fluid" limit.

\subsection{Initial Conditions}

It is important for all codes to start from the same initial condition (or at least as close to one as possible).  This ensures that any distinctions (no matter how small) stem from our numerical methods, and not differences in the true solution, and that all codes would converge to exactly the same solution at sufficiently high resolution.

The initial disk surface density employed here is a simplified version of \citep[][Eq. 4]{MunozLithwick:2020},
%
%\begin{equation}
%    \Sigma(r) = \Sigma_0 \left(\frac{r}{a}\right)^{-\zeta} \left[ 1 - \frac{l_0(q)}{a^2\Omega_b} \right] e^{-\left(r_{\mathrm{cav}/r}\right)^{\xi}}
%\end{equation}
%
\begin{equation}\label{eq:surfaceDensity}
    \Sigma(r) = \Sigma_0 \left[ (1-\delta_0) e^{-(R_{\mathrm{cav} }/r)^{12}}  + \delta_0 \right] f(r)
\end{equation}
with $R_{\rm cav} = 2.5a$ and $\delta_0 = 10^{-5}$ as grid-based codes may have difficulty with such a deep vacuum in their initial condition (this is not a density "floor"; the cavity can get deeper than this during the calculation).

The function $f(r)$ provides a means for truncating the disk, for those codes that are not able to specify general outer boundary conditions.  For an infinite disk, $f=1.$  For a finite disk,
\begin{equation}
    f(r) = 1 - \frac{1}{1+e^{-2(r-R_{\rm out})/a}}
    \label{eqn:fr}
\end{equation}
with $R_{\rm out} = 10a$.  In practice we will find that most results do not depend on whether the finite or infinite disk is chosen (see Section \ref{sec:finite} for a comparison between finite and infinite disk solutions), as previously shown  by \cite{munoz_finite}.

The equation of state is locally isothermal with sound speed
\begin{equation}
    c_s^2 = -\Phi_b(t; r,\phi) h^2 = -\Phi_b(t; r,\phi)/\mathcal{M}^2
\end{equation}
for the (softened) binary potential $\Phi_b = \Phi_1 + \Phi_2$.  We choose a disk with aspect ratio $h = 0.1$ (equivalently an azimuthal Mach number $\mathcal{M} = 10$).

For the initial angular frequency of gas in the disk, we first start with a solution that is in equilibrium far from the binary:

\begin{equation}
    \Omega_0^2(r) = \frac{GM}{r^3}\left( 1 - h^2 \right) = \frac{GM}{r^3}\left( 1 - 1/\mathcal{M}^2 \right)
\end{equation}
suitably modified to flatten in the vicinity of the binary, i.e.
\begin{equation}
    \Omega(r) = \left[ \Omega_0(r)^{-4} + \Omega_B^{-4} \right]^{-1/4}.
\end{equation}

For 3D particle-based runs, we adopted a Gaussian vertical density profile given by
%\extext{-- Equation used by 3D runs. --}
\begin{equation}
    \rho(r,z) = \frac{\Sigma(r)}{H}\frac{1}{\sqrt{2{\rm \pi}}}\exp\left(-\frac{z^2}{2H^2}\right),
\end{equation}
where $H=hr$ is the disk scale height. Note that $\Sigma(r)/H=\rho(r,0)$ is the density on the disk midplane and
\begin{equation}
    \Sigma(r)=\int_{-\infty}^{+\infty}\rho(r,z)dz.
\end{equation}

The 3D \texttt{Athena++} runs adopted a vertically isothermal equation of state, in analogy to the 2D runs, and specified the density profile so as to match the surface density prescribed by Equation (\ref{eq:surfaceDensity}) and to satisfy vertical hydrostatic equilibrium.

We assume a constant kinematic viscosity $\nu$ rather than constant $\alpha$, as different codes may have different $\alpha$ viscosity implementations (this is also consistent with the assumed uniform surface density profile in steady state, far from the binary).  We adopt the value
\begin{equation}
    \nu = 10^{-3} ~a^2 \Omega_B,
\end{equation}
which, for $h = 1/\mathcal{M} = 0.1$, corresponds to $\alpha = 0.1$ at $r=a$, and is also convenient to achieve steady-state solutions in only a few hundred orbits, due to the correspondingly short viscous time in the disk.

A key feature of the isothermal circumbinary disk response to an equal mass binary is an elongated, lopsided ($m=1$) central disk cavity \citep[e.g.,][]{MacFadyen:2008}. In an attempt to seed cavity eccentricity in a uniform way across codes, we implement an initial perturbation or "kick" to the cavity via an initial radial velocity profile of the form
\begin{equation}
    v_r(r) = v_0 ~{\rm sin}(\phi) ~(r/a) ~{\rm exp}\left[ - \left( \frac{r}{3.5a} \right)^6 \right]
    \label{Eq:vrseed}
\end{equation}
with $v_0 = 10^{-4} \Omega_B a$.  This initial perturbation is not included in most binary-disk studies in the literature; is not required for the disk eccentricity growth, but if growth could be seeded by our initial condition rather than numerical noise it might aid agreement between codes. In practice, we will find this initial seed choice is too small to affect the outcome, and particle-based codes that had additional numerical noise unfortunately did not recover the initial value of the disk eccentricity induced by this kick.
We note that we did not give the fluid an initial inward velocity consistent with the viscous drift rate, but this did not make a major difference in the overall solution.

\subsection{Boundary Conditions}

Possibly the biggest difference between numerical setups involves the boundary conditions.  We have tried to design the problem such that differences in boundary conditions would not significantly affect the solution.

\subsubsection{Outer}

The outer boundary is typically chosen to be large enough to not affect the solution (we will test this explicitly in section \ref{sec:finite}).  Because different codes have very different boundaries (e.g. square vs circular, or no boundary at all in the case of particle-based meshless codes)  we do not require a specific outer boundary condition.  \texttt{Disco}, for example, employs a Dirichlet boundary condition at the outer boundary, set at $r=30a$, where the hydrodynamic quantities are fixed by the initial conditions.

\subsubsection{Inner}
The inner boundary condition depends on the setup.  For excised binaries, this is determined at the excision radius ($r=a$).  The standard boundary condition employed for excised binaries will be a "diode" boundary condition, which is essentially a Neumann boundary condition, unless the radial velocity of the fluid is positive, in which case the radial velocity is set to zero to prevent flow of gas from within the excision radius back into the domain.  While most of the non-excised codes resolved the origin at $r=0$, the \texttt{PLUTO} code utilized a grid extending down to a radius of $r = 0.03a$ with a reflecting inner boundary.

\subsection{Sink Prescription}

For non-excised binaries, there is no explicit inner boundary condition, but codes typically require some prescription for accreting mass and momentum ("sink prescription").  This is usually not done using a formal boundary condition, but instead via some prescription for either removing gas particles from the domain, or a sink term explicitly included on the right hand side of the evolution equations for the gas.  This (sometimes loosely-defined) nature of sink prescriptions, can pose a challenge to convergence studies.

For practical reasons, we did not attempt to employ the same sink prescription across different codes.  However, all codes have some prescription for accretion near the individual point masses.  After discussing our fiducial model, we will also assess the effects of different sink prescriptions, to determine their importance, and whether any one sink is a more accurate or well-behaved choice than others.

Particle-based codes \texttt{Gizmo} and \texttt{Phantom} employed sink particles, whereby any fluid parcel that came within a certain distance of the sink (and met certain velocity criteria) was instantly removed from the calculation (and in the case of a live binary, its mass and momentum are added to the sink particle).  Grid-based and moving-mesh techniques typically employed source terms instead; \texttt{Athena++} \texttt{Disco}, \texttt{Mara3}, \texttt{Sailfish} and \texttt{Pluto} all employed some version of this source term, which took the form of an additional expression on the right-hand side of the equations for mass and momentum density evolution.  This source term is then ideally well-resolved in both space and time, such that the size of the affected area is at least a few computational zones across, and the timescale gas is removed on constitutes a large number of timesteps.  Employing a well-resolved sink prescription is generally expected to improve convergence properties of the code.  \texttt{Arepo} employs a sink prescription that is very similar to the grid-based codes, but instead of a slow source term that is resolved in time, it reduces the density towards a fiducial value every timestep, effectively giving a sink timescale that is proportional to the timestep size.

\texttt{Athena++} and \texttt{Disco} each employ different versions of a "torque-free" sink \cite{2020ApJ...892L..29D}.  Roughly speaking, the difference between a "standard sink" and a "torque-free" sink is the choice of how much momentum to remove when removing mass from the system.  We will describe the method in more detail in Section \ref{sec:sinks}, but the idea of a torque-free sink is to effectively be a sub-grid model for what happens on much smaller scales than the sink, assuming the disk extends down to much smaller radii.  Any sink choice will naturally have some effect on the solution at radii outside the sink, but the torque-free sinks were designed to have as little effect as possible, when comparing a disk that is resolved at both large and small radii, to a disk where gas is removed by the sink at some finite radius.

Which sink should be considered "most accurate"?  This depends on the problem one wishes to solve, and what is the physical interpretation for the sink.  If the sink is meant to describe a finite radius where gas is accreted (such as the innermost stable circular orbit of a black hole) then short-timescale "standard" sinks might be a good choice, as they will capture the fact that spin is being added to the point mass.  However, if one wishes to model a system where the inner radius of the disk is meant to be on much smaller scales than that resolved by the code, torque-free sinks are precisely designed and calibrated for this purpose.  Thus, the most accurate choice of sink depends on the physical scenario, and code users should be mindful of the implicit choices being made in the choice of sink prescription.

\section{Diagnostics}\label{sec:diagnostics}

In this section we discuss the different diagnostics we used to compare codes. For each diagnostic the results from each code will be presented as a function of the spatial resolution. We note that, for particle-based/Lagrangian codes, the resolution changes throughout the duration of the simulation. The details about how the resolution was attributed for each diagnostic will be described in the corresponding plot caption.

\subsection{Torque}

In this study, when we describe "gravitational torque" we typically mean the transfer of angular momentum via the gravitational force between the disk and the binary, computed as $\vec T = \vec r \times \vec F_g$, with $F_g = -\int (\nabla \Phi) \Sigma dA$.  For a circular orbit, this is just the radius of the binary component times the $\phi$-component of the gravitational force:
\begin{equation}
    T = r_1 F_1^{\phi} + r_2 F_2^{\phi}
    \label{eqn:torque}
\end{equation}

\begin{equation}
    \vec F_i = \int \Sigma (- \vec \nabla \Phi_i) dA.
\end{equation}
The sign of the torque is defined so that a positive torque corresponds to angular momentum being {\em received} by the binary.  Thus, positive torque works to increase the binary angular momentum (hence separation) and negative torque works to shrink the binary orbit.

In our initial set of runs, we did not compute "accretion torque" or the angular momentum directly accreted onto the binary.  In the limit of a sufficiently small sink radius, or more generally sink terms which do not contribute to the spin angular momentum of either binary component, the orbital angular momentum accreted by the binary is $\dot{M}_1 j_1 + \dot{M}_2 j_2$ (where $M_i$ and $j_i$ are the accretion rate onto and specific angular momentum of binary member $i$), which for an equal-mass binary is by symmetry $0.25 ~\dot{M} a^2\Omega_b$. 

To take into account the non-negligble sink radius, we later found it useful to compute the angular momentum removed by the sink.  Unless otherwise specified, "gravitational torque" computed in this study only accounts for angular momentum transferred directly via gravity, as opposed to "accretion torque".

Often in this study, the torques will be "normalized" in units of $\Sigma_0 a^4 \Omega_B^2$, or $\Sigma_0 GMa$.  However, we have found that the most robust measurement of gravitational torque is obtained by normalizing it to the accretion rate, using the dimensionless quantity $T_{\rm grav}/(\dot M a^2 \Omega_B)$.  This can be related to the accretion "eigenvalue" $l_0$ via
\begin{equation}
    l_0 = \frac{T_{\rm grav} + \dot J_{\rm acc}}{\dot M a^2 \Omega_B},
\end{equation}
where $\dot J_{\rm acc}$ is the angular momentum directly removed by the sinks.   In the limit of small sink radii, $\dot J_{\rm acc} \rightarrow 0.25 \dot M a^2 \Omega_B$ for an equal-mass binary, but we found it useful to compute the accreted angular momentum explicitly; as we will show later, this will result in better agreement between codes.

\subsection{Cavity Eccentricity}

Another important diagnostic to measure is the overall morphology of the disk solution.  In the first 100 orbits of binary evolution, the cavity becomes unstable, breaking the initial symmetry of the system and developing an eccentricity \citep[e.g.,][]{MacFadyen:2008}.  We wish to include a diagnostic that can measure the growth and saturation of this symmetry-breaking instability.  We tested a few diagnostics, and found that the eccentricity vector
\begin{equation}
   \vec e=\frac{|\mathbf{v}|^2\mathbf{r}-(\mathbf{v}\cdot\mathbf{r})\mathbf{v}}{GM}-\hat{\mathbf{r}} \label{eq:e_vector}
\end{equation}
provides the most reliable measurement of the asymmetry of the system.

The $x$ and $y$ components of this vector are
\begin{align}
    e_x &= \frac{r v_r v_\phi}{GM} \sin(\phi) + \left(\frac{r v_\phi^2}{GM}-1 \right)\cos(\phi) \\
    e_y &= -\frac{r v_r v_\phi}{GM} \cos(\phi) + \left(\frac{r v_\phi^2}{GM}-1 \right)\sin(\phi) .
\end{align}
Or, in Cartesian coordinates,
\begin{align}
    e_x &= \frac{j v_y}{GM} - x/r \\
    e_y &= -\frac{j v_x}{GM} - y/r,
\end{align}
with $j = x v_y - y v_x$.  Under this definition, 
\begin{equation}
    |\vec e|=\sqrt{e_x^2+e_y^2}
\end{equation}
is the orbital eccentricity, while
$\varpi$ is the longitude of pericenter such that
\begin{align}
    e_x &= |\vec e|\cos{\varpi} \\
    e_y &= |\vec e|\sin{\varpi}.
\end{align}

We computed a mass-weighted eccentricity vector between $r=a$ and $r=6a$:
\begin{equation}
    \left< \vec e \right> =  \frac{ \int_{a}^{6a} \int_{0}^{2\pi} \vec e(r,\phi) \Sigma dA }{ 35 \pi \Sigma_0 a^2 },
    \label{eqn:eccvec}
\end{equation}
where the denominator is a normalization constant that is designed to reduce this integral to an eccentricity value less than $1$.  The profile of $e(r)$ is another potentially useful diagnostic, however it would require development of a uniform angle-averaging and time-averaging procedure across all codes, which we found cumbersome for a code comparison of this size.

\subsection{Accretion}

Accretion rates are measured directly by asking how much mass was removed by the sinks over the course of a timestep at individual points in time.  These accretion rates are then time-averaged according to the prescriptions given below.

\subsection{Time Averaging and Derivatives}
\label{sec:timeavg}

At sufficiently late times, diagnostics at single points in time become less useful, as there may be a great deal of temporal variability (and this variability is also quite sensitive to the sink prescription, which we are not keeping consistent between codes).  Therefore, we need a process to time-average quantities.  The simplest method would be to take a rolling average or "boxcar" smoothing.  However, such a prescription needs very long duration in order to smooth out variability over a few-orbit timescale.  Ultimately, this is due to the fact that a "top hat" smoothing kernel has power at high frequencies.  An alternative is to use a convolution integral, in particular with a gaussian smoothing kernel:
\begin{equation}
    F_{\rm avg}(t) = \frac{1}{\sqrt{2 \pi} \sigma} \int dt' F(t') e^{-\frac12 (t-t')^2/\sigma^2}.
\end{equation}
By smoothing on a timescale of only $\sigma = 2$ orbits, one can significantly suppress short-timescale variability in most codes, making it possible to focus on long-term trends.  We perform such convolutions over the total torque, excised torque ($r>a$), and the accretion rate, to compare code performance at late times.

It will often be useful to take the derivative of a highly variable function with time (especially the eccentricity vector).  We can similarly define our derivatives by differentiating the smoothed version of a function with time:
\begin{eqnarray}
    \frac{dF}{dt} &\equiv& \frac{dF_{\rm avg}}{dt} = \\ \nonumber
    &=&  \frac{1}{\sqrt{2 \pi} \sigma^3} \int dt'  F(t') (t - t') e^{-\frac12 (t-t')^2/\sigma^2}.
\end{eqnarray}
Thus, our derivatives can also be defined via a convolution integral.  This helps to prevent any one data point from skewing the derivative.

\subsection{Periodograms}

Temporal variability in the accretion rate is decomposed into Fourier modes to generate periodograms.  Similarly to the smoothing procedures above, periodograms are computed over a window of time by multiplying the accretion rate by a Gaussian smoothing kernel:
\begin{equation}
    c(\omega) = \frac{1}{\sqrt{2 \pi} \sigma} \int dt' \dot M(t') e^{-\frac12 (t' - t_0)^2 /\sigma^2} e^{i \omega t'},
    \label{eqn:fourier}
\end{equation}
where $\sigma = 15$ orbits and $t_0 = 250$ orbits.  The power is then computed via $|c(\omega)|^2$.  This means that the accretion variability is measured in a window of time roughly between $200 < t < 300$, but without the high-frequency artifacts that would be present if we cut off the integrals abruptly within the window.  The choice of $\sigma = 15$ orbits means our periodograms are only valid on timescales shorter than this.

\begin{figure}
\includegraphics[width=3.3in]{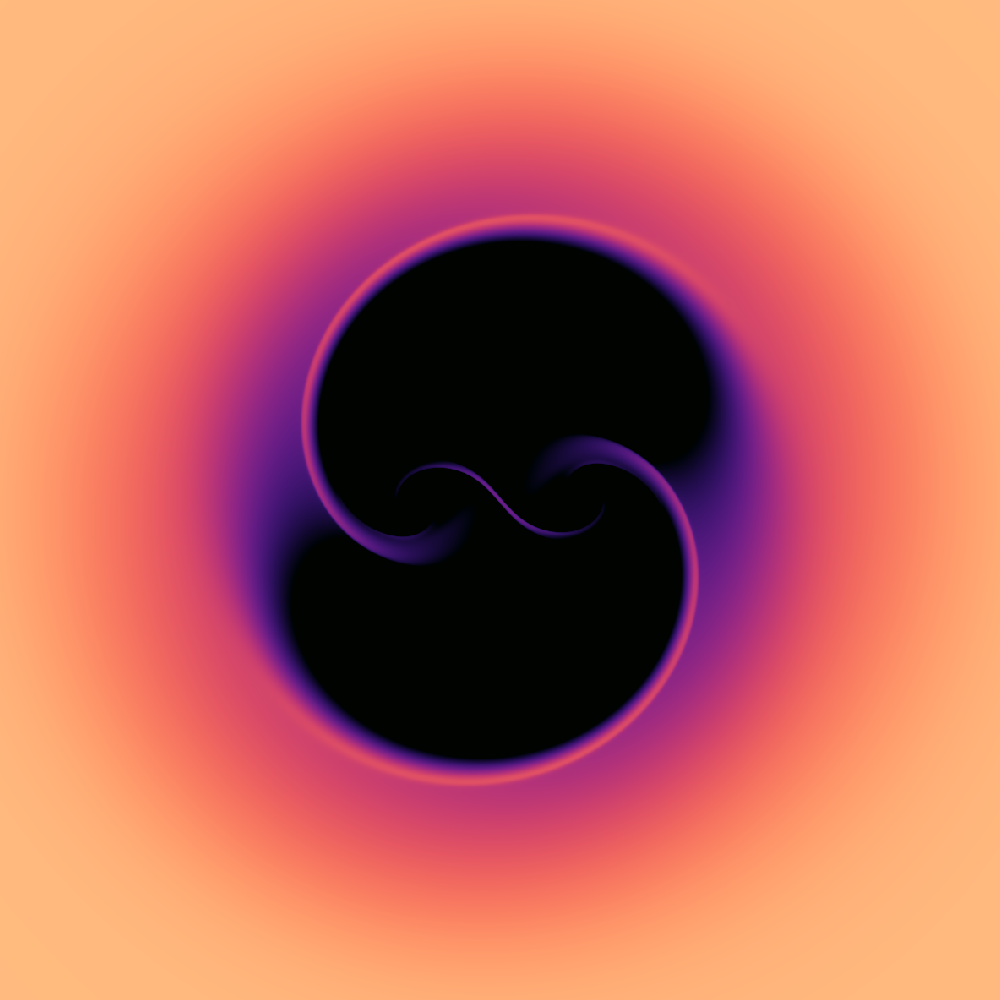}
\caption{Logarithm of the Surface density in the \texttt{Sailfish} code after 1 orbit, at the highest uniform resolution of all codes in the comparison study ($4000 \times 4000$, or $\Delta x = 0.005a$).  For this and other 2D density maps throughout, the spatial grid ranges from $-2.5a$ to $2.5a$ in each dimension, and the colormap is linear, ranging from -3 (black) to 0.5 (yellow).  The colormap is the same as in Figure \ref{fig:1orbit_allcodes}.}
%The colormap is the same as in Figure \ref{fig:1orbit_allcodes}.}
\label{fig:1orbit_hires_001}
\end{figure}

\section{Time Evolution}

We investigate the evolution of the disk for up to 300 binary orbits (with some codes running to 1000 orbits), by which time the system has settled into a relaxed (quasi) steady-state.  Before this time, the solution still depends on the details of the initial conditions; we explore the evolution from this initial state to the final state at different stages.  We first look in detail at the first orbit, in part as a consistency check on the details of the initial conditions.

\begin{figure}
\includegraphics[width=3in]{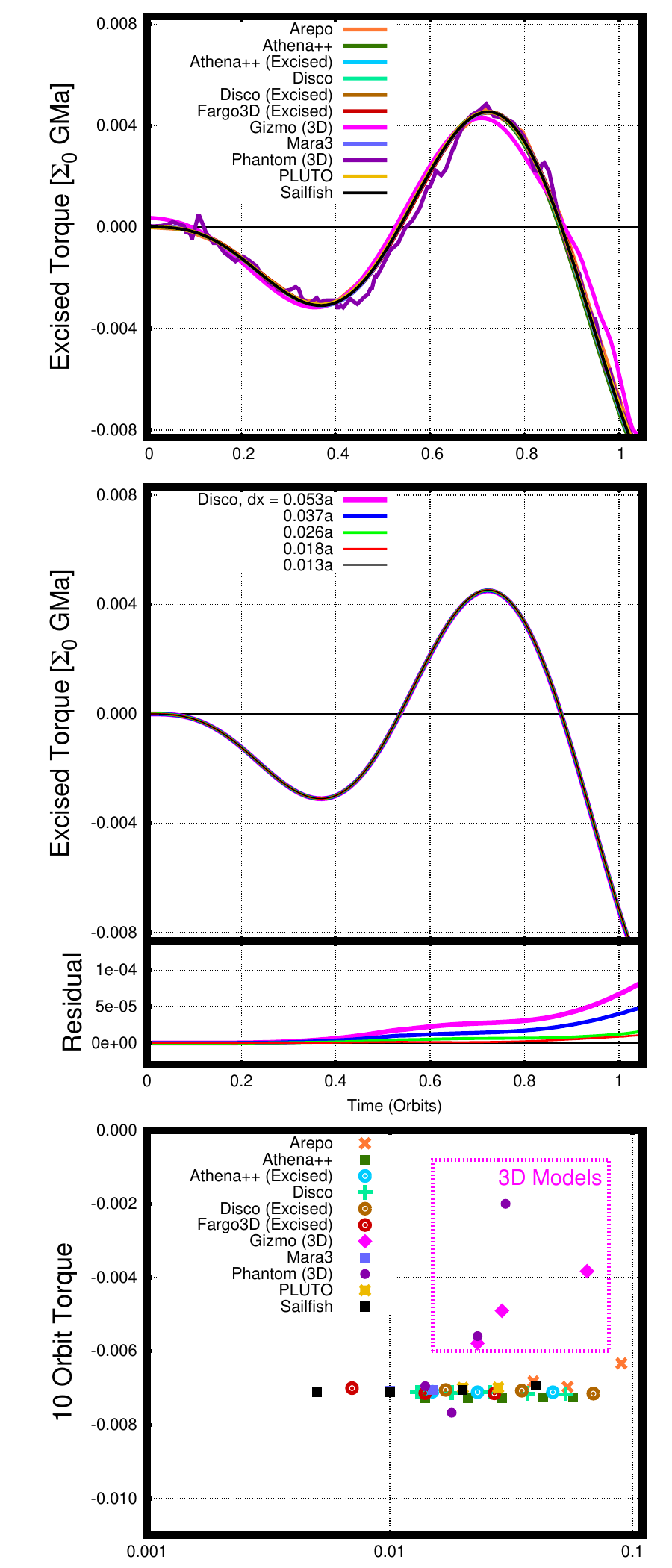}
\caption{Torque convergence over the first orbit.  All codes are plotted in the top panel, lying on essentially the same curve (consistent with the highest-resolution \texttt{Sailfish} run).  A demonstration of convergence in the \texttt{Disco} code is shown in the center panel, showing very little deviation between resolutions.  Residuals are included (taking the difference betwween the output and the highest-resolution run), showing rapid convergence of the solution.  The value of the torque at one binary orbit is plotted in the lower panel, for all codes, as a function of the resolution length scale.  The resolution is measured in terms of the zone size $\Delta x/a$ at $r = 3a$.}
\label{fig:1orbit_torq}
\end{figure}

\begin{figure*}
\includegraphics[width=7.1in]{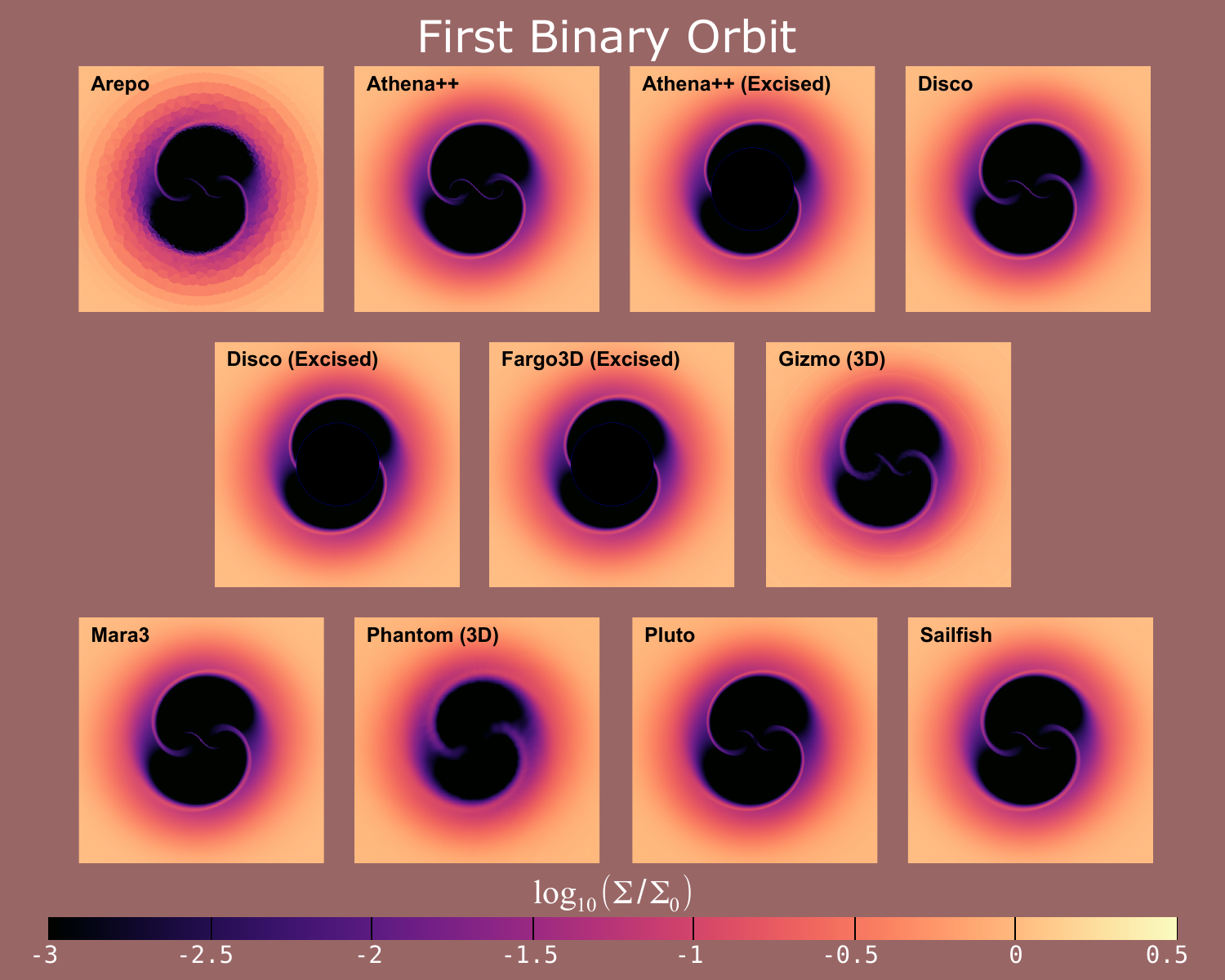}
\caption{Surface density for all codes at 1 orbit.  Codes are compared with similar zone spacing $\Delta x = 0.02a$ at $r = 3a$.  During this first orbit, tenuous streams of gas begin to accrete onto the binary and extremely low-density minidisks begin to form. Part of each stream passes by the first binary member encountered leading to a higher-density shock near the barycenter.} 
\label{fig:1orbit_allcodes}
\end{figure*}

\subsection{The First Orbit}

Figure \ref{fig:1orbit_hires_001} shows the high-resolution results obtained using the \texttt{Sailfish} code after 1 orbit, as an example of the calculations performed.  During the first binary orbit, the initially axisymmetric disk begins to viscously spread and is strongly perturbed by the gravitational influence of the binary. Nascent circumprimary and circumsecondary accretion disks (generically `minidisks') form, initially from the ambient gas within the cavity, but later gas stripped from the circumbinary disk. Some of the gas that is stripped from the spreading cavity walls streams past the binary member which initially perturbed it and shocks against the gas symmetrically perturbed by the other binary member, leading to a higher-density bridge-like feature between them.

The first orbit provides an opportunity to test that all codes have implemented the setup and initial conditions in precisely the same way.  Capturing the evolution of the first orbit is reasonably straightforward; in this first orbit, all codes quickly converge to the same solution.  Figure \ref{fig:1orbit_torq} shows the torque as a function of time during this first orbit, again showing that a converged torque is found and agreed-upon by all codes at these times.  Excised torque was computed (using contributions from fluid elements with $r>a$) in part because all codes could be compared, including those that excised the binary.  However, during the first orbit (and similarly the first ten orbits), very little torque comes from within $r<a$, as very little gas has accreted onto the binary at this stage.

The center panel of Figure \ref{fig:1orbit_torq} shows convergence of the \texttt{Disco} code at five different resolutions, demonstrating that the solution is very well converged.  Increasing the resolution by a factor of four has almost no effect on the curve $T(t)$; differences between the different resolution runs are at the percent level.

The lower panel of Figure \ref{fig:1orbit_torq} shows the excised torque at one orbit as a function of resolution for all codes.  All codes converge toward the same value of this torque.  The converged excised torque at one orbit is found to be $T/(\Sigma_0 G M a) = -7.2 \times 10^{-3} \pm 10^{-4}$, where error bars are roughly set by differences between codes.  The only (minor) deviations are from particle-based codes that have additional Poisson noise from the positions of the initial computational particles (or zones).  These codes can still capture time-averaged torques reliably at these times, as seen in the next section.

Figure \ref{fig:1orbit_allcodes} shows a plot of surface density at one orbit for all numerical methods in our study.  In this and future large comparative images, all codes are using the same effective resolution of $\Delta x = 0.02a$ at $r=3a$ (though some codes have higher resolution on the minidisks and lower resolution further out in the disk).  The basic structures are captured similarly by all codes; particle-based code \texttt{Phantom} exhibits less well-defined shocks.  We later demonstrate this to be a resolution-dependent effect, which can be cured by very high particle count; because particle-based codes typically resolve mass elements rather than volume elements, at these early times when no gas is yet in the cavity they can have very low effective spatial resolution near the origin (though in this study \texttt{Gizmo} mitigates this effect via particle splitting, see \cite{Franchini2022} for details).

\begin{figure}
\includegraphics[width=3.3in]{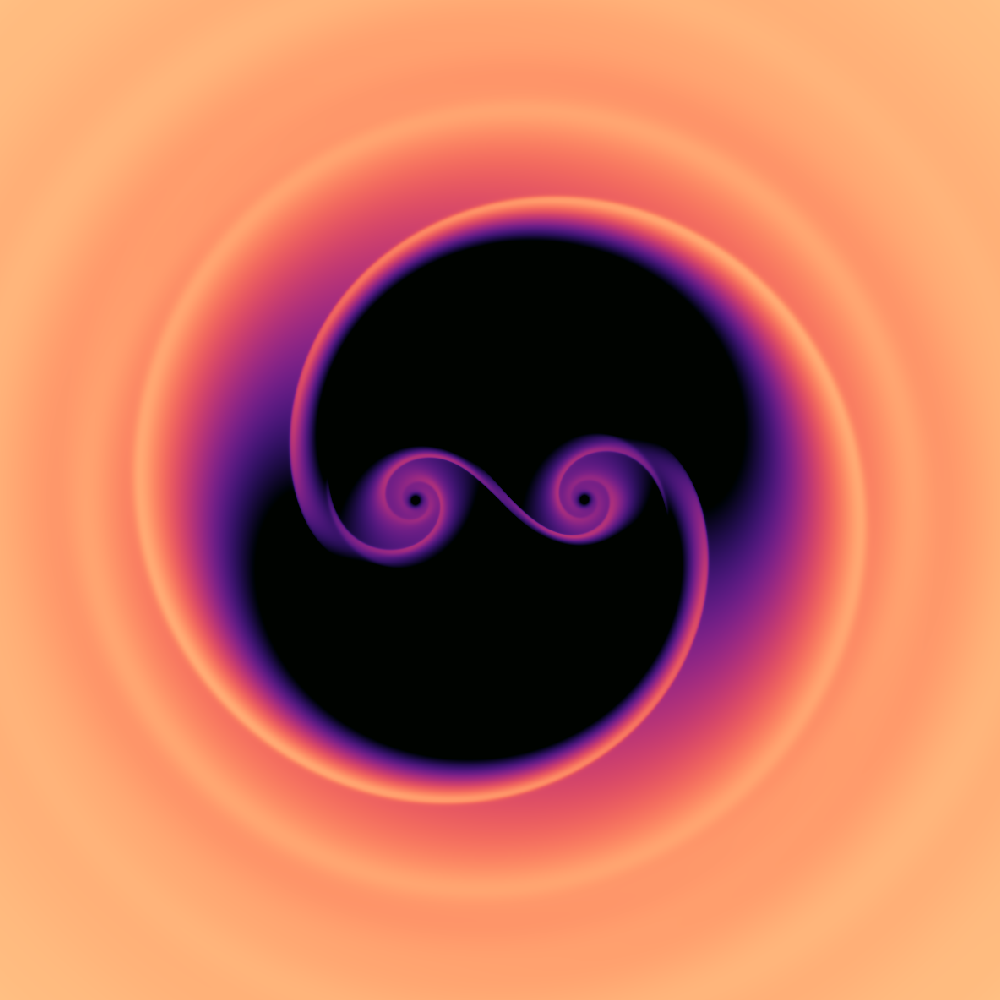}
\caption{Logarithm of the Surface density in the \texttt{Sailfish} code after 10 orbits. The colormap is the same as in Figure \ref{fig:10orbit_allcodes}.}
\label{fig:1orbit_hires_010}
\end{figure}

\subsection{The First 10 Orbits}

%\begin{figure}
%\includegraphics[width=3.3in]{conv_10orbit.pdf}
%\caption{Torque convergence in the first ten orbits.  The excised torque ($r>a$) from the Sailfish code is compared at different resolutions, showing that the code approaches a consistent torque as a function of time at sufficiently high resolution.  For Sailfish, one requires at least $dx \lesssim 0.02a$ for a reasonable torque.} 
%\label{fig:conv_10}
%\end{figure}

\begin{figure}
\includegraphics[width=3.3in]{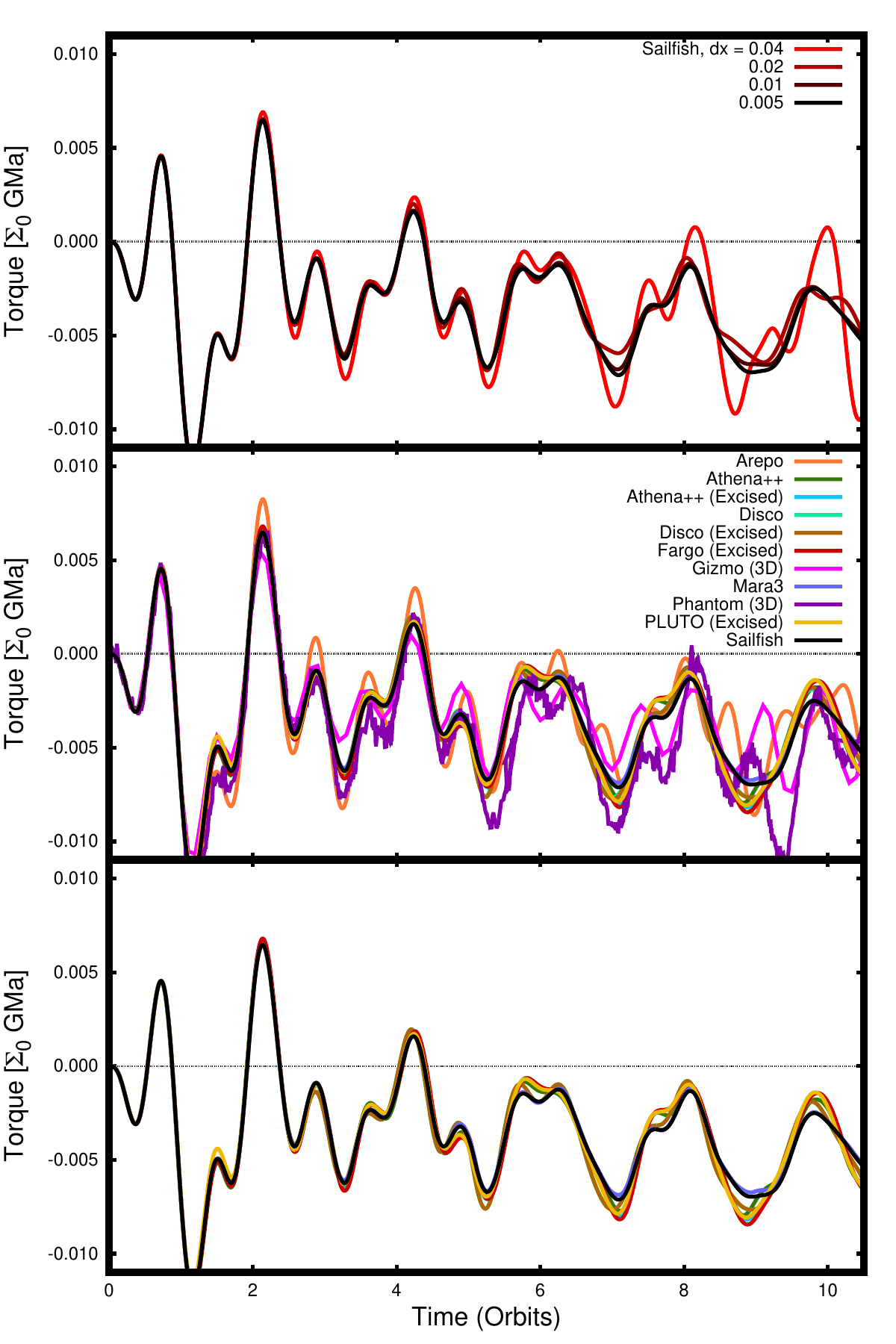}
\caption{Torque comparison between all codes in the first ten orbits.  In the top panel, the excised torque ($r>a$) from the \texttt{Sailfish} code is compared at different resolutions, showing that the code approaches a consistent torque as a function of time at sufficiently high resolution.  All codes are plotted at their highest resolution in the center panel; the lower panel highlights a subset of eight codes which agree the most precisely during the first 10 orbits.} 
\label{fig:10orbit_torq}
\end{figure}

\begin{figure}
\includegraphics[width=3.3in]{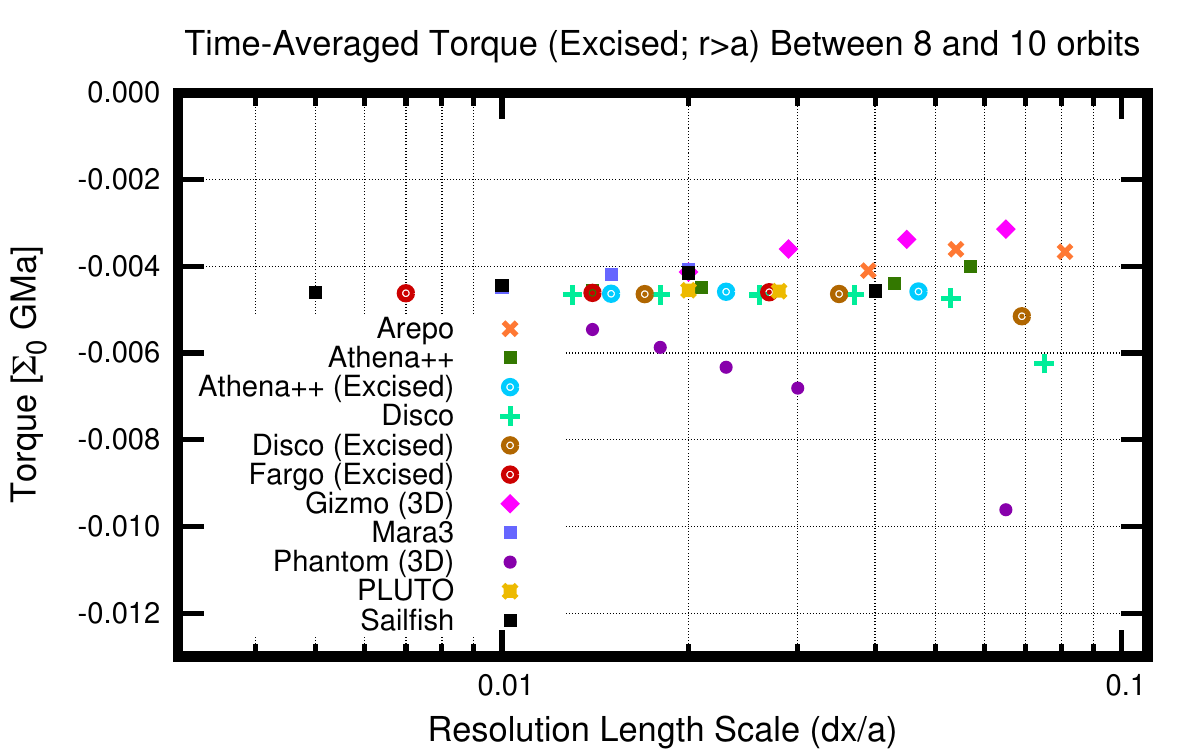}
\caption{Convergence of the time-averaged torque from 8-10 orbits for all codes in this study.} 
\label{fig:conv_10}
\end{figure}

\begin{figure*}
\includegraphics[width=7.2in]{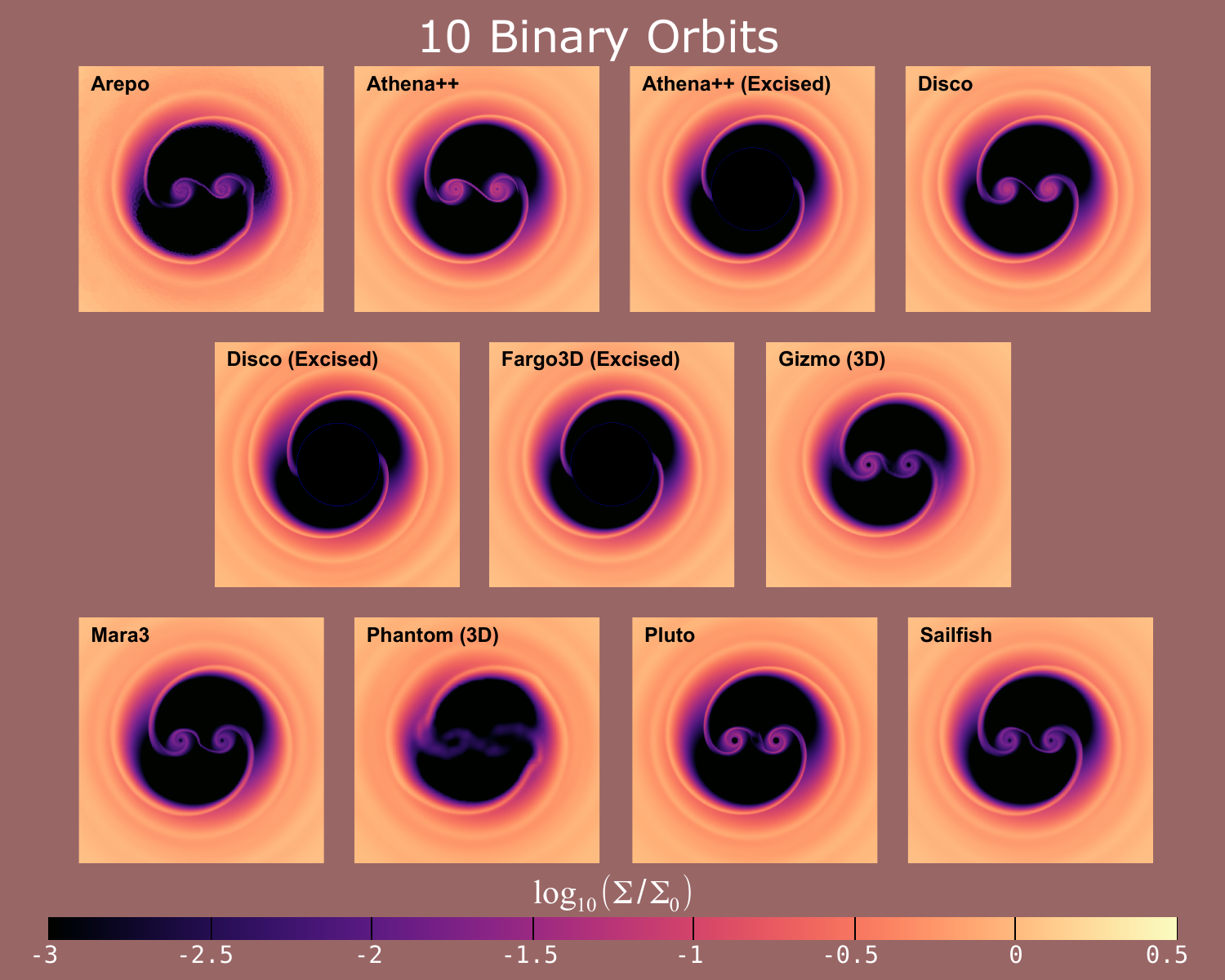}
\caption{Surface density for all codes at 10 orbits.  Codes are compared with similar zone spacing $\Delta x = 0.02a$ at $r = 3a$.  Although discrepancies exist between codes at this resolution, it should be noted that many of the codes are capable of running at much higher resolution than in this figure; the idea is to attempt to compare all codes roughly the same zone or particle spacing at $r=3a$.  Note that this means some codes will still have much better resolution near the origin.} 
\label{fig:10orbit_allcodes}
\end{figure*}

\begin{figure}
\includegraphics[width=1.65in]{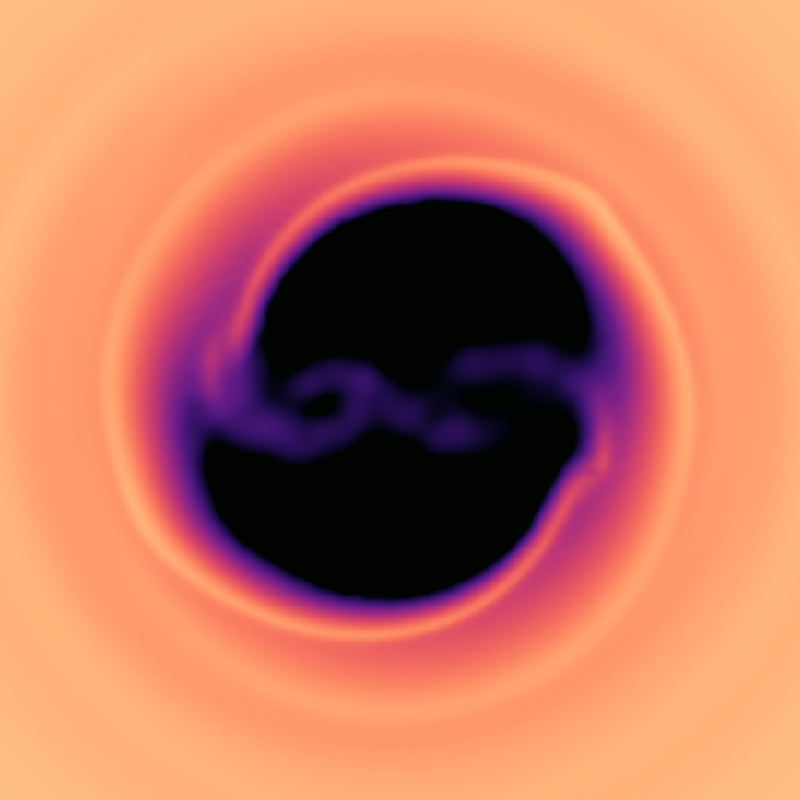}
\includegraphics[width=1.65in]{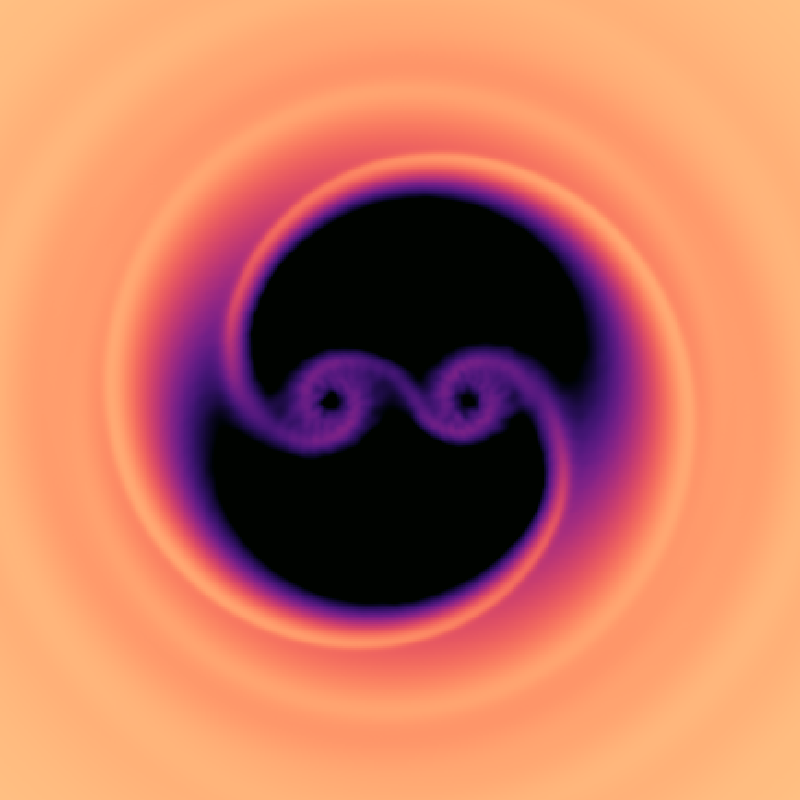}
\caption{\texttt{Phantom} code at 10 orbits, comparing resolutions of 20 million particles (left) vs 100 million particles (right).  The colormap is the same as in Figure \ref{fig:1orbit_allcodes}.}
\label{fig:10orbit_ragusa}
\end{figure}

After the first few orbits, the accretion of gas onto the binary settles into a fairly regular pattern, as some gas from the disk is captured by the binary forming minidisks.  Waves have begun to propagate outward through the circumbinary disk, and some gas is able to flow from one minidisk to the other.  An example showing the calculation after ten orbits at high resolution is shown in Figure \ref{fig:1orbit_hires_010} (using the \texttt{Sailfish} code as an example).  On a ten-orbit timescale, nonlinear effects make convergence less trivial, but still reasonably attainable.  The minidisks have begun to accumulate a bit of gas already, but nearly all torque still comes from outside $r>a$.  The cavity still appears to be symmetric, respecting the symmetry of the binary and the initial conditions.  Figure \ref{fig:10orbit_torq} shows the torque as a function of time in the \texttt{Sailfish} code at many resolutions.  It is evident that \texttt{Sailfish} is able to achieve a converged torque as a function of time at sufficiently high resolution (at least for these first 10 orbits).

The central panel of Figure \ref{fig:10orbit_torq} compares all of the codes during this time, using the lower panel to highlight which codes attain very similar torque curves.  While the outliers exhibit somewhat noisier torque curves, all codes are able to agree on the time-averaged torque over these timescales; Figure \ref{fig:conv_10} shows the time-average of the torque between 8 and 10 orbits, computed by a simple numerical integration over this window of time.  All codes at all resolutions are plotted, demonstrating that it is possible for all codes to achieve the converged solution at sufficiently high resolution.  During the first ten orbits, the gravitational torque is usually negative.  Between 8 and 10 orbits, the time-averaged gravitational torque is $T/(\Sigma_0 G M a) = -0.0046 \pm 10^{-4}$, where the uncertainty is (roughly) quantified using the spread between different codes in the plot at their highest resolutions.

Figure \ref{fig:10orbit_allcodes} shows the surface density at ten orbits for all numerical methods in our study.  As the minidisks accumulate mass, the particle-based codes show improved spatial resolution within the cavity, but discrepancies in minidisk morphology still remain, specifically when considering the \texttt{Phantom} code.  In figure \ref{fig:10orbit_ragusa}, we demonstrate that \texttt{Phantom} can achieve agreement with the other codes, when the cavity is sufficiently well-resolved.  However, at 10 orbits the particle count in the cavity is still sufficiently low in the fiducial runs that the minidisks are yet unresolved.  Interestingly, each code appears to find a different minidisk surface density at this time, likely reflecting discrepancies in sink prescriptions between codes.  Excised codes are able to capture the outer disk morphology accurately.

Figure \ref{fig:10orbit_allcodes} indicates that binary-disk interaction is a challenging problem. After only 10 orbits, significant visual discrepancies in minidisk morphology develop across the different codes.  This visual trend will continue at later stages.  However, when measuring averaged quantities such as torque or accretion rate, we will find that the different codes can reach agreement.  Thus, despite visual differences in the instantaneous snapshots, it will be possible to find convergence to a common value for bulk integrated quantities.

\subsection{The First 100 Orbits}

\begin{figure}
\includegraphics[width=3.3in]{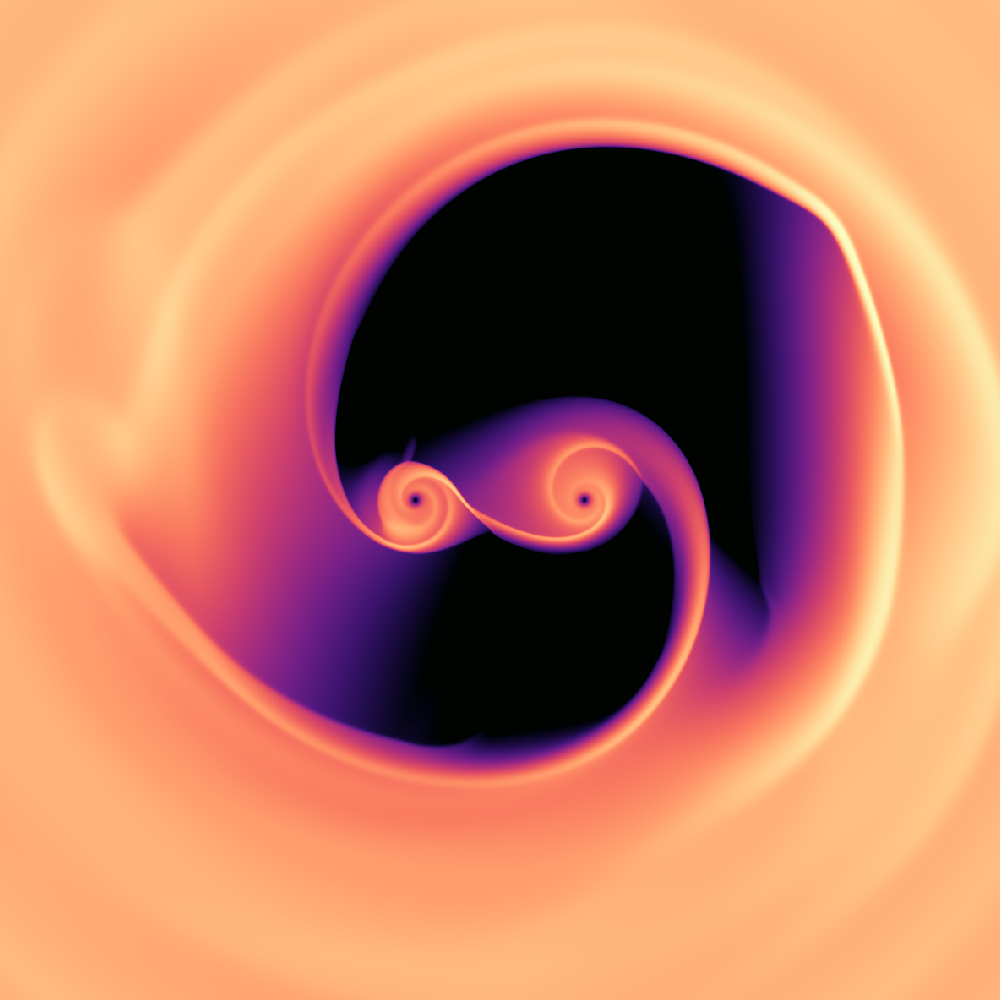}
\caption{Logarithm of the Surface density in the \texttt{Sailfish} code after 100 orbits. The colormap is the same as in Figure \ref{fig:100orbit_allcodes}.}
\label{fig:1orbit_hires_100}
\end{figure}

\begin{figure}
\includegraphics[width=3.3in]{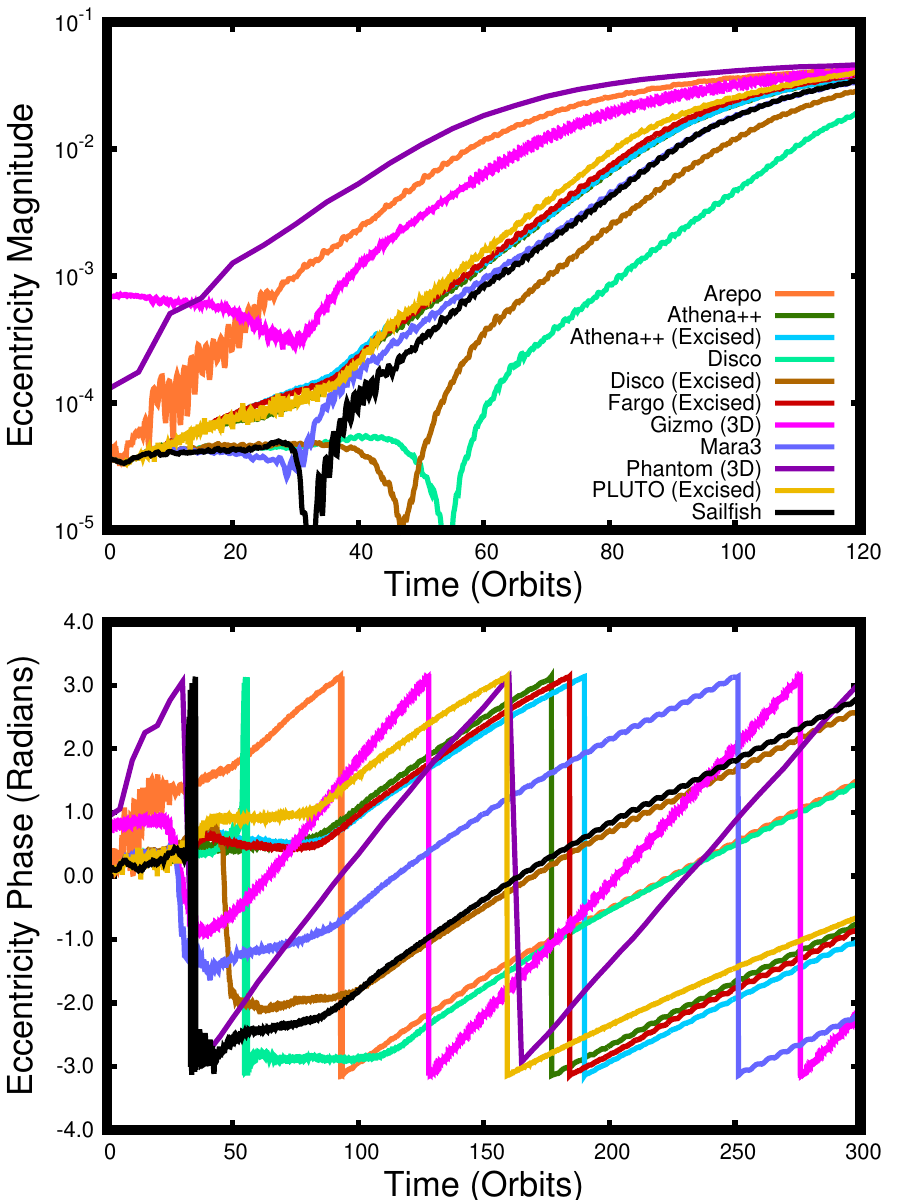}
\caption{Eccentricity vector over time for all codes.  The top panel shows the magnitude and the bottom panel shows the phase.  The growth of the instability and subsequent precession of the cavity begin at different times for different codes.  Nevertheless, the growth rate and precession rate have reasonably consistent values when comparing between codes.} 
\label{fig:ecc_time}
\end{figure}

\begin{figure}
\includegraphics[width=3.3in]{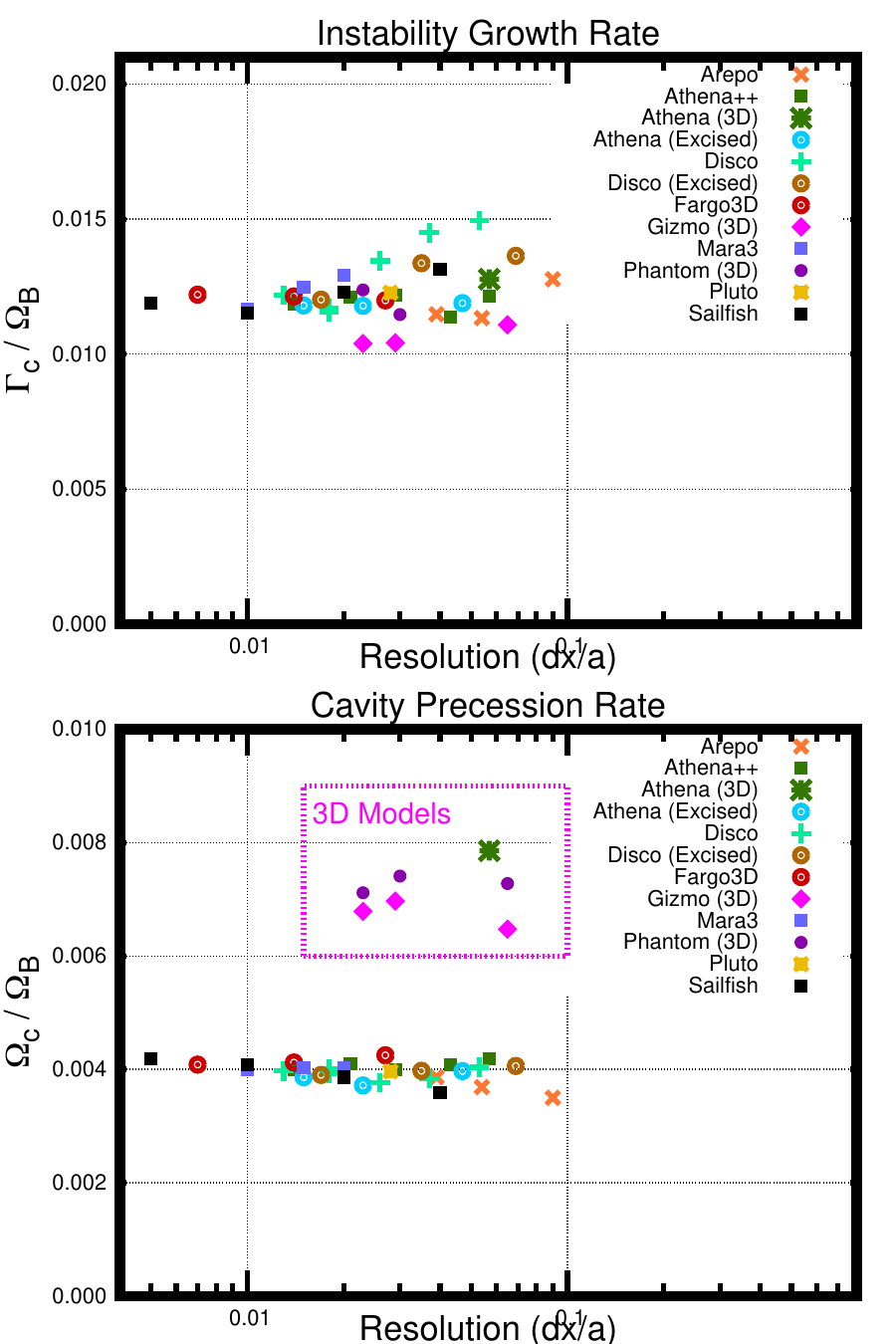}
\caption{Growth rate and cavity precession rate for all codes at all resolutions.  The precession rate is measured at 150 orbits.  The growth rate is measured at a time when the eccentricity has grown to $e = 3\times10^{-3}$, when all codes find exponential growth (this occurs at 
different times for different codes, see Figure \ref{fig:ecc_time}).  The growth rate ($\Gamma_c/\Omega_B$) converges to a rate of about 0.012, corresponding to an e-folding time of about 13 binary orbits.  The cavity precession period ($2\pi/\Omega_c$) converges to a period of about 240 binary orbits.  3D codes find a shorter period, about 150 binary orbits.} 
\label{fig:ecc_res}
\end{figure}

\begin{figure*}
\includegraphics[width=7.1in]{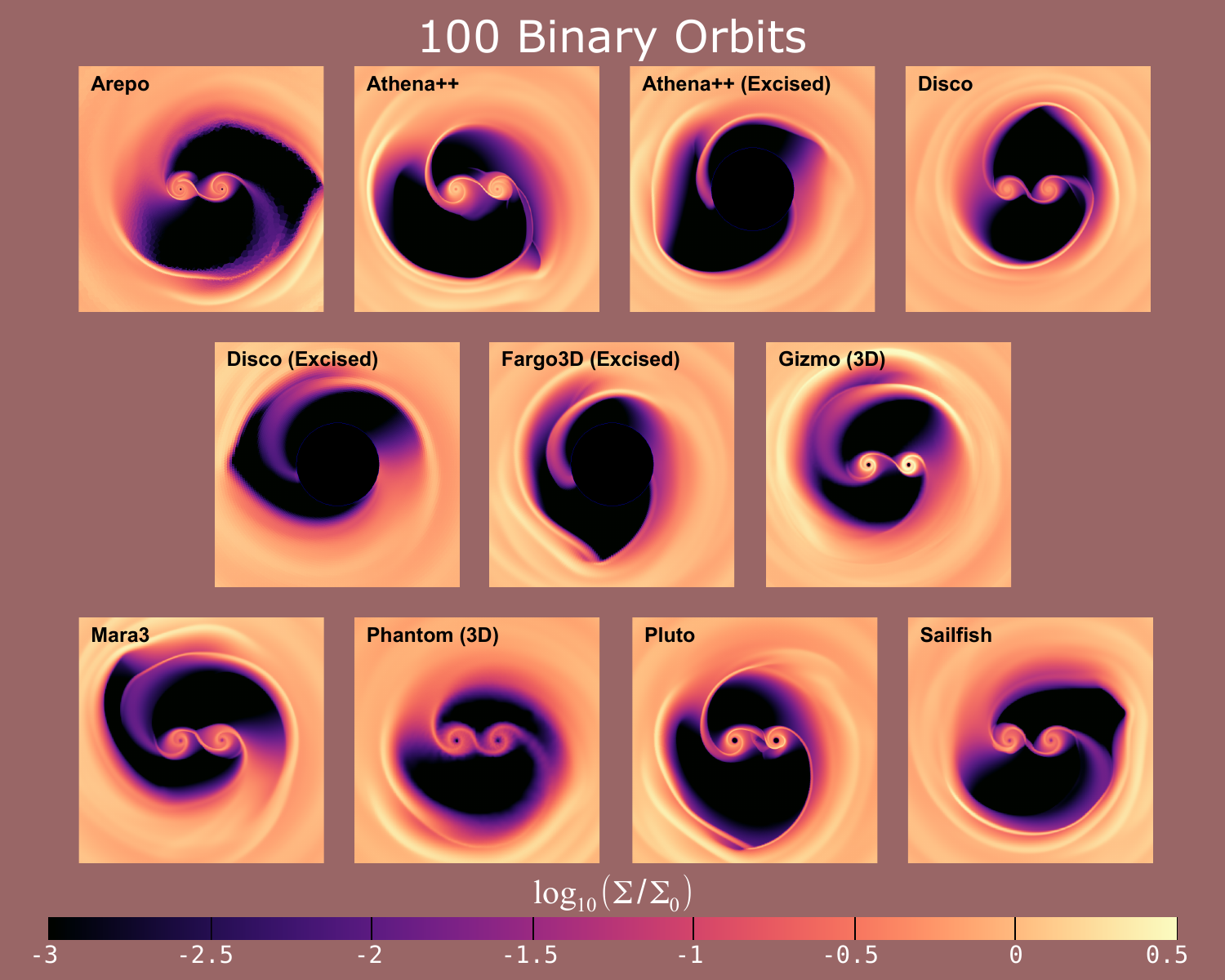}
\caption{Surface density for all codes at 100 orbits.  Codes are compared with similar zone spacing $\Delta x = 0.02a$ at $r = 3a$.  The discrepancies in surface density between codes result from the different rates and times at which disk eccentricity grows in each simulation, as illustrated in Figure \ref{fig:ecc_time}. Though we seeded this instability with a finite perturbation, the resulting growth is very sensitive to details, and results in a different measured cavity phase between codes.  In addition, the cavity precession rate is about 1.6 times as fast in 3D, so that there should be no expected agreement in the orbital phase of the asymmetry between 3D and 2D.} 
\label{fig:100orbit_allcodes}
\end{figure*}

\begin{figure}
\includegraphics[width=3.3in]{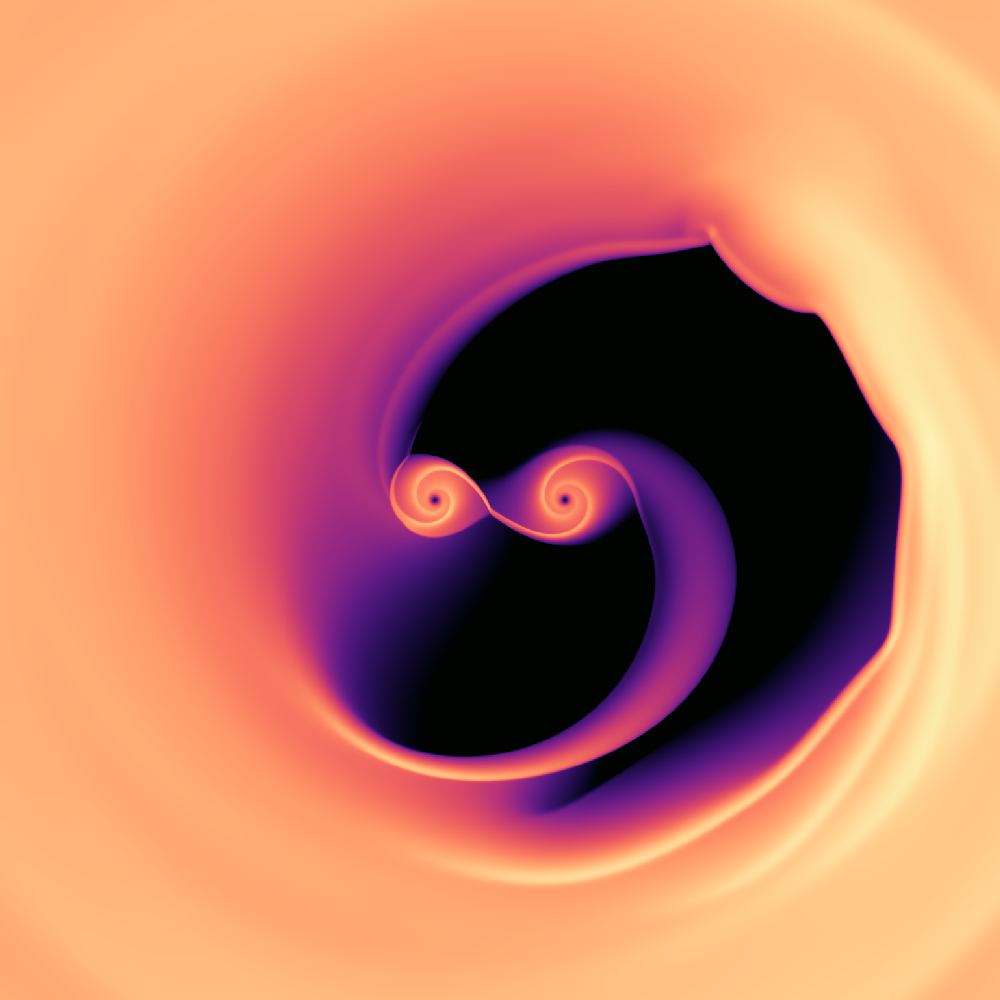}
\caption{Logarithm of the Surface density in the \texttt{Sailfish} code after 300 orbits. The colormap is the same as in Figure \ref{fig:300orbit_allcodes}.}
\label{fig:1orbit_hires_300}
\end{figure}

An example showing the calculation after 100 orbits at high resolution is shown in Figure \ref{fig:1orbit_hires_100} (using the \texttt{Sailfish} code as an example).  Until this point, the disk has been nearly symmetric.  During the first hundred orbits, however, this picture gradually changes. This eccentric ``instability'' was first noted in \cite{MacFadyen:2008}, which used the \texttt{FLASH} code with an excised binary.  It has been characterized by \cite{ShiKrolik:2012, DHM:2013:MNRAS} and has been shown to manifest itself at mass ratios as low as $q\gtrsim0.01$, although its amplitude varies depending on disk and binary properties \citep[e.g.,][]{D'Orazio:CBDTrans:2016, ragusa2020, noble21, dittmann_q}. The conditions for its stability and growth rate as a function of all disk and binary parameters have not yet been completely explored.

A model of the instability was given in \cite{ShiKrolik:2012}.  Streams carrying momentum and mass are excited by the binary and impart energy and momentum onto fluid elements at the cavity edge. A preference for one stream to becoming stronger than the other, creates a feedback cycle which causes stream impacts on one side of the cavity to dominate \citep{DHM:2013:MNRAS}, hence increasing the eccentricity of the inner circumbinary disk and cavity. An overdense "lump" builds up and orbits at the cavity edge. This lump orbits with the orbital period of the cavity edge, which for the 2D simulations in this work is very close to five binary orbital periods. However, size of the cavity, and thus its orbital period, depend strongly on the properties of both the binary and the disk \citep[e.g.,][]{DHM:2013:MNRAS,Farris:2014,noble21, dittmann_q}, so this five-orbit periodicity is far from universal. 

Eccentricity growth impacts all measured quantities.  For example, prior to eccentricity saturation, all codes found a negative torque on the binary, which would lead to inward migration (for this test problem), and a sudden shift in the torque towards outward migration occurs around the time when the instability nears saturation.  This was true for all codes, regardless of when the cavity eccentricity saturated.  The two appear to be causally linked, but the details depend on the disk and binary parameters. In some cases, disks that are stable and do not achieve an eccentric cavity morphology result in inward migration \citep{Franchini2022, wang23, dittmann_q}.

We have found that the growth of this instability and subsequent precession of the cavity are clearly represented using the density-weighted eccentricity vector $\vec e$ defined previously (\ref{eqn:eccvec}).  The magnitude of this vector over time captures the growth and saturation of the cavity eccentricity, while the phase of this vector nicely tracks the precession of the cavity.

We plot each of these quantities as a function of time in Figure \ref{fig:ecc_time}.  All codes find similar growth and saturation of the instability, though some codes find growth beginning at different start times from others.  3D codes notably find a faster cavity precession rate, by roughly $\sim60\%$.

The growth rate $\Gamma_c$ can be measured as $\Gamma_c = \dot e / e$:

\begin{equation}
    \Gamma_c \equiv \frac{\dot e_x e_x + \dot e_y e_y }{ e_x^2 + e_y^2 }.
\end{equation}

The precession rate $\Omega_c = \partial_t (\varpi)$, or

\begin{equation}
    \Omega_c = \frac{\dot e_y e_x - \dot e_x e_y }{ e_x^2 + e_y^2 }.
\end{equation}

$\Gamma_c$ and $\Omega_c$ are computed for all codes, at all resolutions, and the result is plotted in Figure \ref{fig:ecc_res}.  The final saturated value of the eccentricity depended on whether the disk was finite or infinite, and on whether the binary was excised, and on whether the disk was modeled in 2D or 3D.  Thus, there was no consensus on a final saturated value of $e$, even though the converged growth rate was consistent between all codes.

All codes show an exponential behaviour, with a growth rate  $\Gamma_c/\Omega_B \approx 0.012$ (about a 13-orbit e-folding timescale) that appears roughly consistent across all codes (in both 3D and 2D). The cavity precession rate is $\Omega_c/\Omega_B \approx 0.004$ (240 orbit precession period) for all the 2D codes.  In 3D, the cavity precesses 60\% faster, giving a period of about 150 binary orbits.  To disambiguate between effects of code choice and effects of 3D vs 2D, we have also run \texttt{Athena++} in 3D and \texttt{Gizmo} in 2D.  We will discuss further the differences between 2D and 3D in Section \ref{sec:dep}.

Figure \ref{fig:100orbit_allcodes} shows a plot of surface density at 100 orbits for all numerical methods in our study.  At this time, it can be difficult to compare these images directly, although this is a natural result of the nonlinear hydrodynamic processes that we have simulated. Because disk eccentricity was seeded by numerical perturbations, each disk began developing eccentricity at a different time and with a slightly different preferred orientation, leading to a variety of cavity orientations after and leading up to saturation.  

\subsection{Relaxed Disk -- 300 Orbits}

\begin{figure}
\includegraphics[width=3.3in]{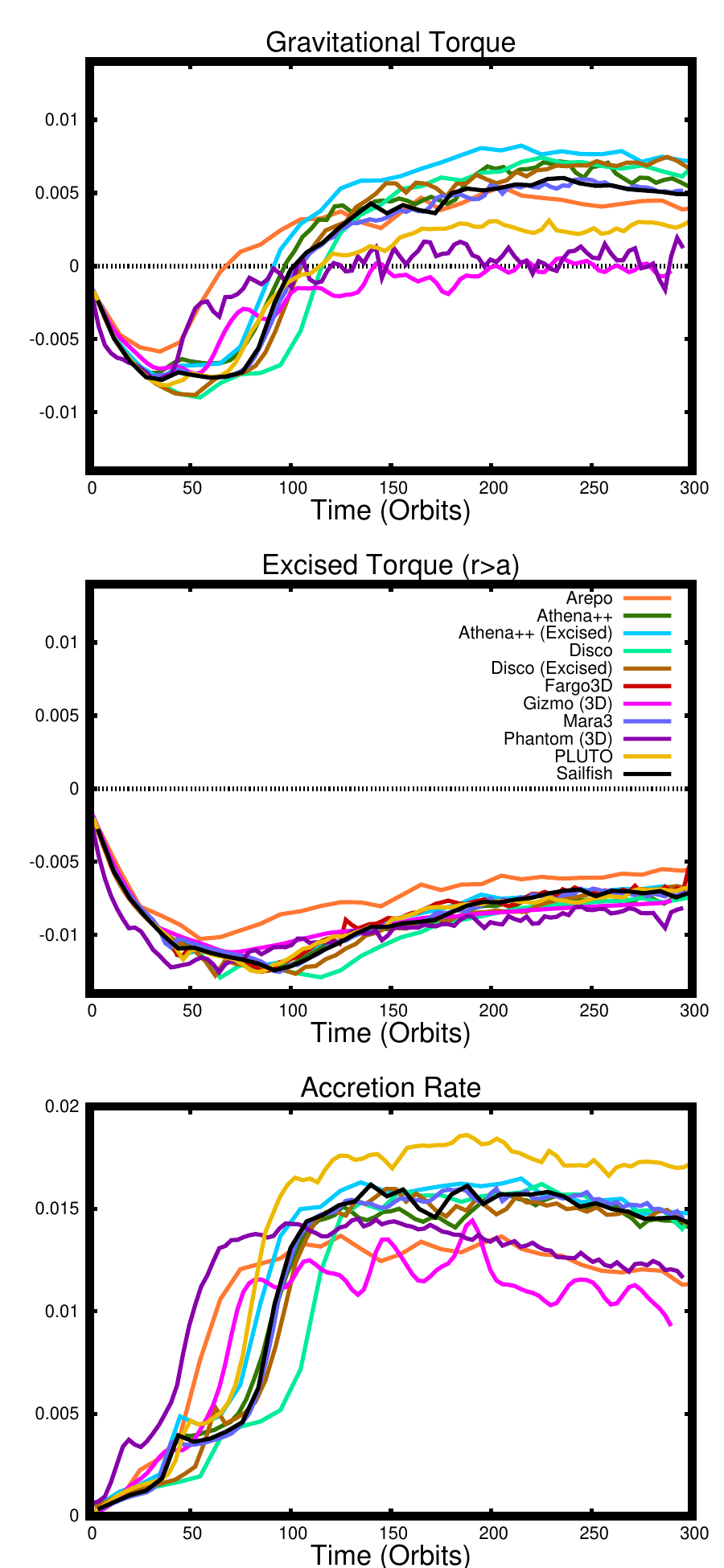}
\caption{Smoothed torque and accretion rates as a function of time for all codes in this study.  At around 100 orbits, this torque flips sign from negative to positive, demonstrating the importance of running codes for sufficiently long timescales to allow the instability to develop in the disk.  The center panel shows the excised torque, from fluid elements outside $r>a$.  The bottom panel shows the smoothed accretion rate.  The largest discrepancies in accretion rate can be traced back to discrepancies in the initial disk models (finite vs. infinite disk, see section \ref{sec:finite}).} 
\label{fig:timeavg_torq}
\end{figure}

The maximum time that most of the codes were run for was 300 orbits.  As a representative example, Sailfish is shown at high resolution in Figure \ref{fig:1orbit_hires_300}.  300 orbits is sufficient for disk eccentricity to saturate and the disk to attain a quasi-steady state (this is confirmed in the next subsection, where some of the codes are run to 1000 orbits). After the disks become visibly eccentric, codes are unable to agree at specific moments in time. This is because the growth of the instability is sensitive to small details.  Though we tried to mitigate this somewhat by including an initial seed perturbation (Eq. \ref{Eq:vrseed}), these details nonetheless offset the phase of the cavity relative to the binary at late times, and therefore comparing codes at a single time becomes less meaningful.  We therefore require time-averaging in order to mitigate this effect and extract long-timescale trends that can be compared between codes.  To this end, we perform a convolution integral with a Gaussian kernel (details in section \ref{sec:timeavg})  to time-average the values of torque and accretion rate as a function of time, with the results reported in Figure \ref{fig:timeavg_torq}.

\begin{figure}
\includegraphics[width=3.3in]{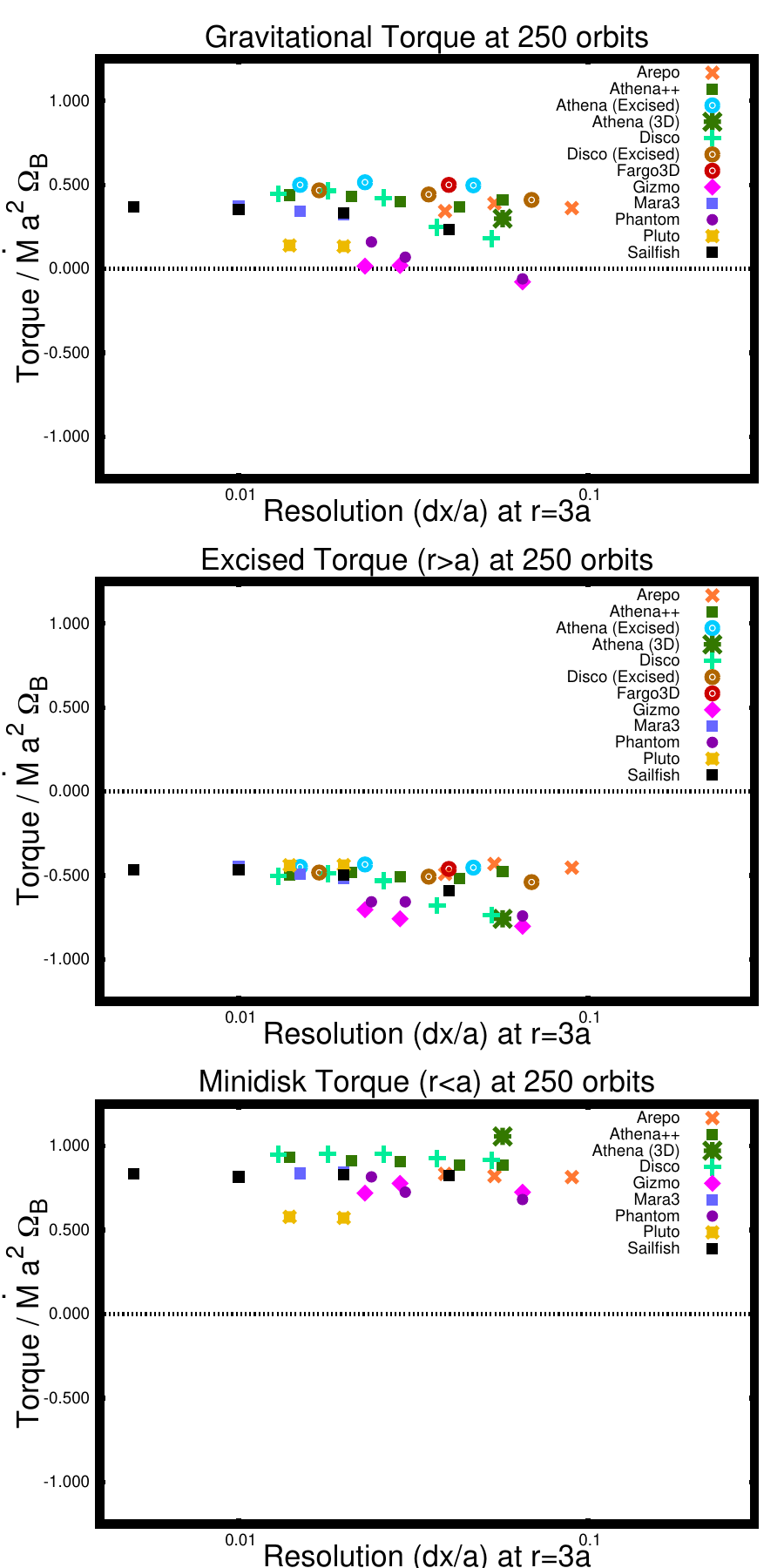}
\caption{Time-averaged gravitational torque, normalized by $\dot M a^2 \Omega_B$, for all codes at all resolutions.  The top panel shows the total gravitational torque on the binary, whereas the center and bottom panels separate this into excised torque (from $r>a$) and "minidisk torque" (from $r<a$).  On the surface it appears that the different codes disagree, but these discrepancies can be resolved by considering the dimensionality of the problem and adjusting for the sink prescription (seen in section \ref{sec:sinks}).} 
\label{fig:300torq}
\end{figure}

Torques switch sign around 100 binary orbits -- coincident with the growth and saturation of the cavity eccentricity.  In Figure \ref{fig:300torq}, we plot the gravitational torque (normalized to accretion rate) for all codes in this study, as a function of resolution.  Here we find significant discrepancies between our codes, so it will now be necessary to diagnose these discrepancies and determine whether they are code-related or due to minor differences in our code setup.  At first glance it appears there is a cluster of codes that find torques of $T_{\rm grav} \sim 0.4-0.5 \dot M a^2 \Omega_B$, while \texttt{PLUTO}, \texttt{Gizmo} and \texttt{Phantom} all converge to much smaller gravitational torques.  In fact, taken at face value, these three codes would find inward migration while the rest of the codes would predict outward migration.

The largest difference between some of codes is in the dimensionality of the setup; \texttt{Gizmo} and \texttt{Phantom} performed calculations in 3D, whereas the rest of the codes performed the calculation in 2D.  Actually, it is a little surprising that differences between 2D and 3D are not more severe.  We will also show that the next major difference between codes is in the choice of sink prescription, which is responsible from the deviations of \texttt{PLUTO} from the other 2D codes. However, discrepancies in the sink prescription can be accounted for by properly accounting for the angular momentum eaten by the sink.  Thus, while Figure \ref{fig:300torq} appears to show significant discrepancies between codes, we will find that these discrepancies can be understood and resolved (See section \ref{sec:sinks}). 

\begin{figure*}
\includegraphics[width=7.1in]{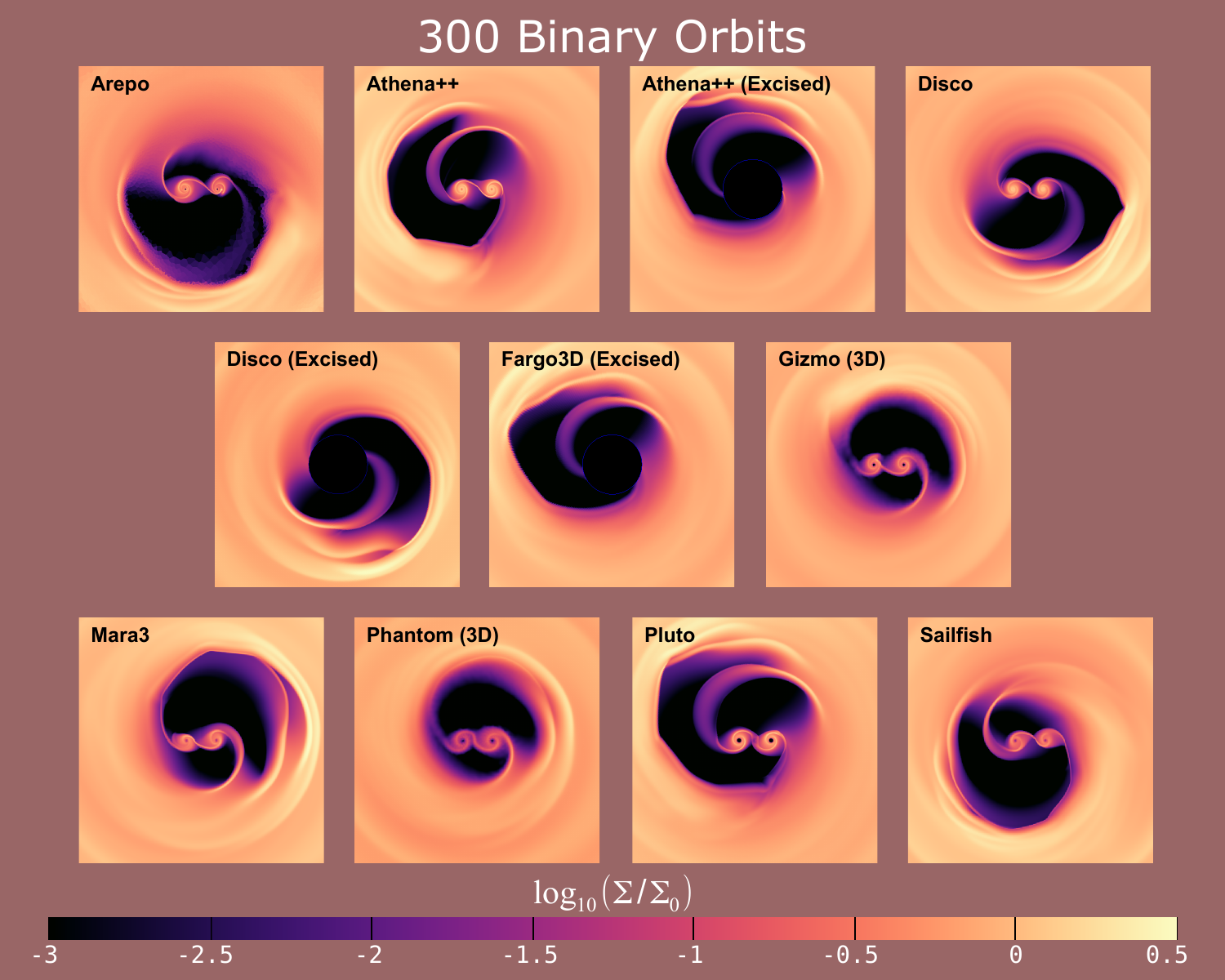}
\caption{Surface density for all codes at 300 orbits.  Codes are compared with similar zone spacing $\Delta x = 0.02a$ at $r = 3a$.  At this stage, the asymmetry and size of the cavity have reached their peak, and we see the saturated state of the instability.  At this point, the cavity precesses and the lump of gas responsible for accretion variability orbits the edge of the cavity, but otherwise the disk has reached a "relaxed" state; still time-dependent and therefore not steady-state, but arguably quasi-steady-state.} 
\label{fig:300orbit_allcodes}
\end{figure*}

\subsection{Accretion Periodograms}

\begin{figure*}
\includegraphics[width=7.1in]{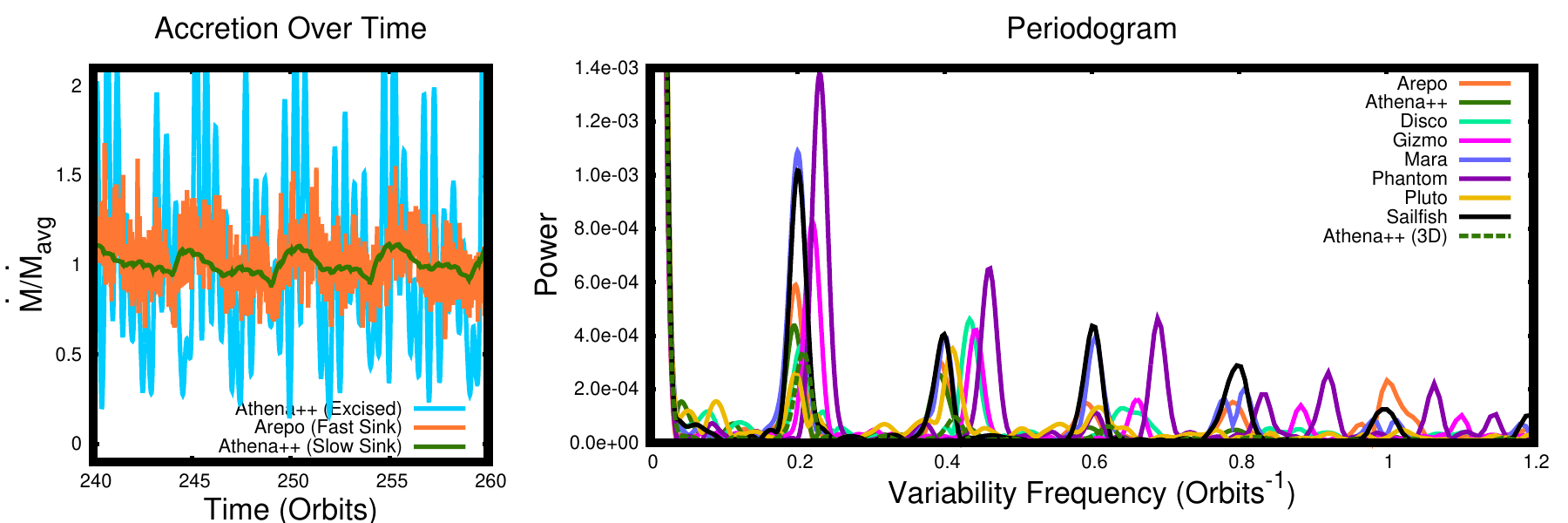}
\caption{Accretion rate periodograms for a subset of the codes in our study.  The choice of sink prescription or boundary condition can alter predictions for variability on different timescales.  This is related to how much minidisk buffering is expected for a given inner truncation radius of the minidisk. Short-timescale variability will depend on this choice; it is possible that eccentric binaries might have variability less dependent on the sink or boundary choice.  The left panel demonstrates what this looks like in the form of a time-series; the excised \texttt{Athena++} run (cyan) was included just to demonstrate the most extreme version of this effect; as the cavity was excised in that run, there can be no minidisk buffering at all.} 
\label{fig:periodogram}
\end{figure*}

As one additional code comparison at late times, we compute accretion periodograms, by taking the Fourier transform of $\dot M(t)$ according to Eq.~\eqref{eqn:fourier} and computing the square of the amplitude $c^2(\omega)$.  This is plotted in Figure \ref{fig:periodogram}.  All codes in our study find a variability peak on both 1 orbit and 4-5 orbit timescales, in addition to many other peaks in the periodogram.  However, the amplitudes of the various peaks in the periodogram show a great deal of variation between codes.  Much of this discrepancy is likely due to the different choices in sink prescription.  We note that \texttt{Athena++} and \texttt{Disco}, for example, exhibit less variability than some of the other codes on 1 orbit and shorter timescales. In contrast, codes such as \texttt{Sailfish} and \texttt{Arepo} find very large variability on orbital timescales and shorter.  Particle-based codes \texttt{Gizmo} and \texttt{Phantom} exhibit even higher degrees of variability on all timescales.

The differences in sink parameters used by each code are responsible for the most major differences in the periodograms between codes, with faster, larger, and standard (vs torque-free) leading to greater variability. \citet{dittmann_sinks} explored this in a more controlled setting, demonstrating that at a constant sink rate "standard" sinks lead to higher-amplitude variability than torque-free sinks, and that for a given sink method faster sinks lead to higher amplitude variability (though to a much smaller extent when using torque-free sinks). The same behaviour - the presence of minidisks reducing (quasi-)periodic accretion signatures - has also been observed on a physical rather than numerical basis in numerical relativity simulations \citep[e.g.,][]{2023MNRAS.520..392B}.  

As to the question of which sink prescription is more accurate, it depends on what the user intends to model on small scales.  The torque-free sinks are designed to model disks that extend down to radii well below the grid scale. However, if the sink is meant to represent a resolved inner radius of the minidisk, such as the ISCO of a black hole, then a faster sink (and one that permitted accretion of spin) might mimic such an inner boundary condition. Ultimately, one must try to match their sink prescription to the physical problem at hand, and keep in mind the effects of these numerical choices on the periodicities identified by each simulation.

\subsection{1000 Orbits -- Checking the Steady-State}

\begin{figure}
\includegraphics[width=3.3in]{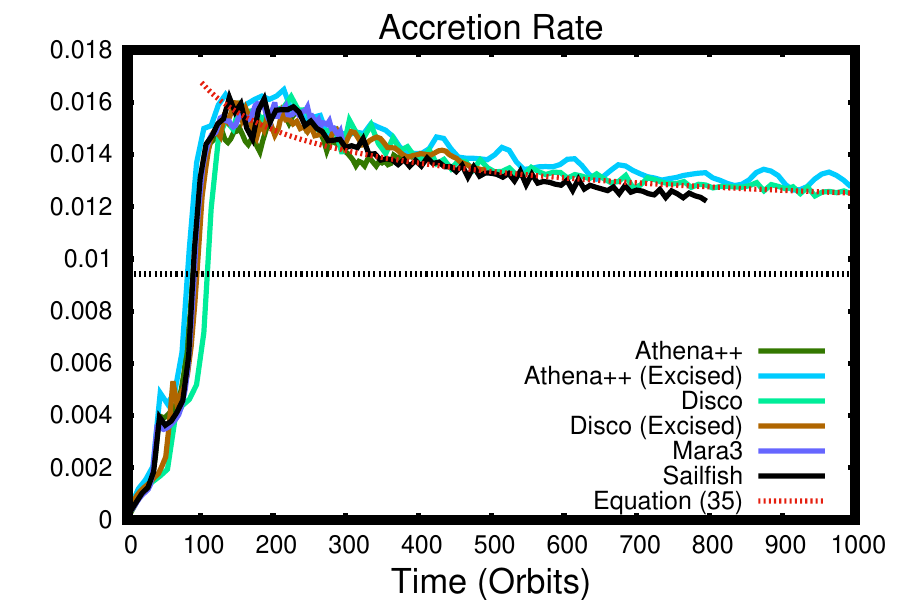}
\caption{Comparison of the accretion rate reported by codes with infinite disks and the asymptotic value for accretion onto a single point mass (dashed grey line).  $\dot M$ matches the analytical expression (\ref{eqn:mdot_asym}), which very slowly asymptotes to the single-mass value after many thousands of orbits.  In this transient period, the accretion rate is enhanced by the presence of the binary.} 
\label{fig:mdot_long}
\end{figure}

By running our codes for a thousand orbits, it becomes possible to establish the time it takes to reach (quasi) steady-state.  The relaxed state of this system is typically achieved after about 200 binary orbits.  This timescale is determined both by the growth rate of the instability and the number of e-foldings that the system must grow by to reach saturation.  Accretion rate as a function of time up to 1000 orbits is plotted in Figure \ref{fig:mdot_long} for those codes which employed an infinite disk model.

Because our initial conditions assumed zero angular momentum current through the disk \citep[c.f.][Equation (\ref{eq:surfaceDensity})]{LBPringle:1974,MunozLithwick:2020} but the angular momentum current through the disk after settling into a quasi-steady state through interaction with the binary was positive ($\dot{J}\approx0.7\dot{M}$), our simulations measured greater accretion rates than the steady-state value ($\dot{M}_0=3\pi\nu\Sigma_0$). This enhancement following underestimation of the angular momentum current was demonstrated in \citet{MirandaLai+2017}, and relative suppression in the case of overestimation of the angular momentum current was noted in \citet{dittmann_mach}. 

However, our simulations do gradually approach the asymptotic quasi-steady accretion rate, given sufficient time.  Since the viscous time at a given radius $R$ is $t_\nu \sim R^2/\nu$, one can define a viscous radius $R_\nu(t)$, the radius within which we have achieved steady-state, and outside of which is out of equilibrium (and determined by initial conditions),
\begin{equation}
    R_\nu(t) \sim (\nu t)^{1/2}.
    \label{eqn:rnu}
\end{equation}
Inside of this radius, the steady-state can be determined by the flow of mass and angular momentum
\begin{equation}
    \dot J = \dot M j - 3 \pi \nu \Sigma j,
\end{equation}
where $\dot J$ and $\dot M$ are both positive quantities, defined as the inward flow of angular momentum and mass, respectively.  The right-hand side of the equation is given by the advective flux of angular momentum (first term) and the outward flow of angular momentum due to viscosity (second term).

This equation can be inverted to determine a solution for $\Sigma(R)$ within $R<R_\nu(t)$,
\begin{equation}
    \Sigma(R) = \frac{\dot M - \dot J/j(R)}{3 \pi \nu},
\end{equation}
where $j(R) = R^2 \Omega(R)$ is the specific angular momentum at the radius $R$.  Using $\dot J = l_0 \dot M a^2 \Omega_B$,  this can be expressed as
\begin{equation}
    \Sigma(R) = \dot M \frac{1 - l_0 \sqrt{a/R} }{3 \pi \nu},
\end{equation}
which is the solution within $R<R_\nu$.  Outside, we can assume the disk has not had time to change its surface density and $\Sigma = \Sigma_0$.  This effectively sets a boundary condition of $\Sigma = \Sigma_0$ at $R = R_\nu(t)$, so that
\begin{equation}
    \Sigma_0 = \dot M \frac{1 - l_0 \sqrt{a/R_{\nu}(t)} }{3 \pi \nu}.
\end{equation}
Now we can invert this expression to find a solution for $\dot M$ as a function of time,
\begin{equation}
    \frac{\dot M(t)}{3\pi \nu \Sigma_0} = \frac{1}{1 - l_0 \sqrt{a/R_\nu(t)}} .
\end{equation}

Using the expression (\ref{eqn:rnu}) above for $R_\nu$ and including a dimensionless overall coefficient $\kappa$,
\begin{equation}
    R_\nu(t) = \kappa \sqrt{\nu t},
\end{equation}
we arrive at the following expression for $\dot M$,
\begin{equation}
    \frac{\dot M(t)}{3\pi \nu \Sigma_0} = \frac{1}{1 - l_0 (\kappa^2 \nu t/a^2)^{-1/4}}.
    \label{eqn:mdot_asym}
\end{equation}
Using the eigenvalue $l_0 = 0.7$, we find the coefficient $\kappa = 3.2$ gives a good fit to the accretion rate (see Figure \ref{fig:mdot_long}).

For an infinite disk, this accretion rate will eventually trend towards the asymptotic value $3 \pi \nu \Sigma_0$ (the value one would find for a single point mass with no angular momentum input), but the $1/4$ power-law in the angular momentum term ensures that the approach to steady-state is extremely slow. Running our codes for about 40 thousand binary orbits would be necessary to achieve a value within $10\%$ of the asymptotic value.  In practical terms, the binary torques can change the resultant disk accretion rate over very long timescales.

To summarize, the time-averaged accretion rate onto the binary asymptotes towards a value identical to the value for a single point mass, but the convergence toward this rate is slow and there can be significant enhancement in the accretion rate (by up to a factor of 2) for many viscous times.  Nevertheless, even though the accretion rate has not leveled out, the disk has reached a "relaxed" state; for example, the ratio of torque to accretion rate has reached a steady constant value, as we will show in Section \ref{sec:finite}.

\section{Dependence on Numerical Choices}
\label{sec:dep}

We have already discussed a few discrepancies in some of the measured diagnostics between codes.  We now assess whether these discrepancies are due to numerical choices.  If so, what controls them, and which (if any) method is more accurate.  Additionally, we would like to test the effects of additional numerical choices, such as the gravitational softening radius, live binary orbit, and other choices implicitly made by our codes.

\subsection{2D vs 3D}

The dimensionality of our disk is arguably not a numerical choice, but a choice of what problem to solve.  Nevertheless, some codes are tailored to run in 3D, and have therefore performed a 3D version of this problem.  Here, we investigate what effect this has on our computed diagnostics.

Our measurements of the gravitational torque show significant discrepancies.  In particular, the particle-based codes \texttt{Gizmo} and \texttt{Phantom} show a significantly lower torque than the other codes (though \texttt{PLUTO} also shows a smaller torque than the others).  Interestingly, \texttt{Gizmo} and \texttt{Phantom} also find a much faster cavity precession rate. Notably, \texttt{Gizmo} and \texttt{Phantom} are solved 3D version of the problem, whereas the other codes are being run in 2D.  So, any discrepancy could easily be explained away by arguing these codes are solving a different problem than the others.  Unfortunately that would not teach us much about how these codes compare with each other.  So, we have performed an experiment where a grid-based code (\texttt{Athena++}) is run in 3D and a particle-based code (\texttt{Gizmo}) is run in 2D, to determine which of these discrepancies can be explained purely by the dimensionality of the system.

We first investigate the cavity precession rate.  We find strong evidence that the precession rate is very different in 3D than in 2D, sufficient to explain entirely the discrepancy seen earlier.  \texttt{Athena++} finds a precession rate of $\Omega_c/\Omega_B \approx 0.008$ (125 orbit period) in 3D, whereas at the same resolution in 2D the precession rate was found to be $\Omega_c/\Omega_B \approx 0.004$, consistent with the converged value in 2D.  So 3D gives a precession rate that is faster by about a factor of 2.  This was similarly tested by running \texttt{Gizmo} in 2D.  In 3D, the converged precession rate was $\Omega_c/\Omega_B \approx 0.007$ for \texttt{Gizmo}.  In 2D, we found a slower precession rate of $\Omega_c/\Omega_B \approx 0.004$, in agreement with the other codes.

The simplest explanation for the difference in precession rate between 2D and 3D follows from the smaller cavity sizes observed in the 3D simulations, which cause the circumbinary disk to experience a more non-Keplerian potential and thus precess more rapidly \citep[e.g.,][]{MunozLithwick:2020}. However, in 3D there is an additional pressure term in the evolution equation for the disk eccentricity, which can potentially provide a comparable contribution to the precession rate \citep[e.g.,][]{2006MNRAS.368.1123G,teyssandier2016}.  This discrepancy in the precession rate between 2D and 3D is likely some combination of the two effects.

A more complicated issue is the torque. \texttt{Gizmo} and \texttt{Phantom} exhibited a significantly lower gravitational torque than the other codes: low enough that one might expect the binary to migrate inwards, rather than outwards. \cite{moody19} (using the grid-based \texttt{Athena++} code) also found smaller torques in 3D, but the discrepancy was not as large; \cite{moody19} predict outward migration both in 2D and 3D.

We can shed some light on this puzzle by separating the torque into two components: an "excised torque" which adds up contributions with $(r>a)$ and a "minidisk torque", i.e. the remaining component, from the minidisks themselves $(r<a)$.  When comparing 2D and 3D torques from \texttt{Athena++} and \texttt{Gizmo}, we find that the {\em excised} torque ($r>a$) is significantly different between 2D and 3D, enough to explain a large portion of the discrepancy between codes. However, the {\em minidisk} torque ($r<a$) does not show a large discrepancy between 2D and 3D, and instead the spread between codes is much smaller (this remaining discrepancy is caused by the differences in sink prescription, see Section \ref{sec:sinks}).  So, we can reasonably confirm that all codes correctly compute a converged excised torque that agree between codes, but there is a different answer in 2D than in 3D, a difference of about $0.16 \dot M a^2 \Omega_B$, which was noted in \citealt{moody19}.

We find a remaining significant discrepancy in the minidisk torque, but here there is no major dividing line between codes -- as seen in the lower panel of Figure \ref{fig:300torq}, every code seems to converge to a different minidisk torque, with a general spread of about $0.5 \dot M a^2 \Omega_B$.  Our next step is to investigate whether this remaining discrepancy could be caused by discrepancies in our sink prescriptions.

\subsection{Sink Prescription}
\label{sec:sinks}

\begin{figure}
\includegraphics[width=3.3in]{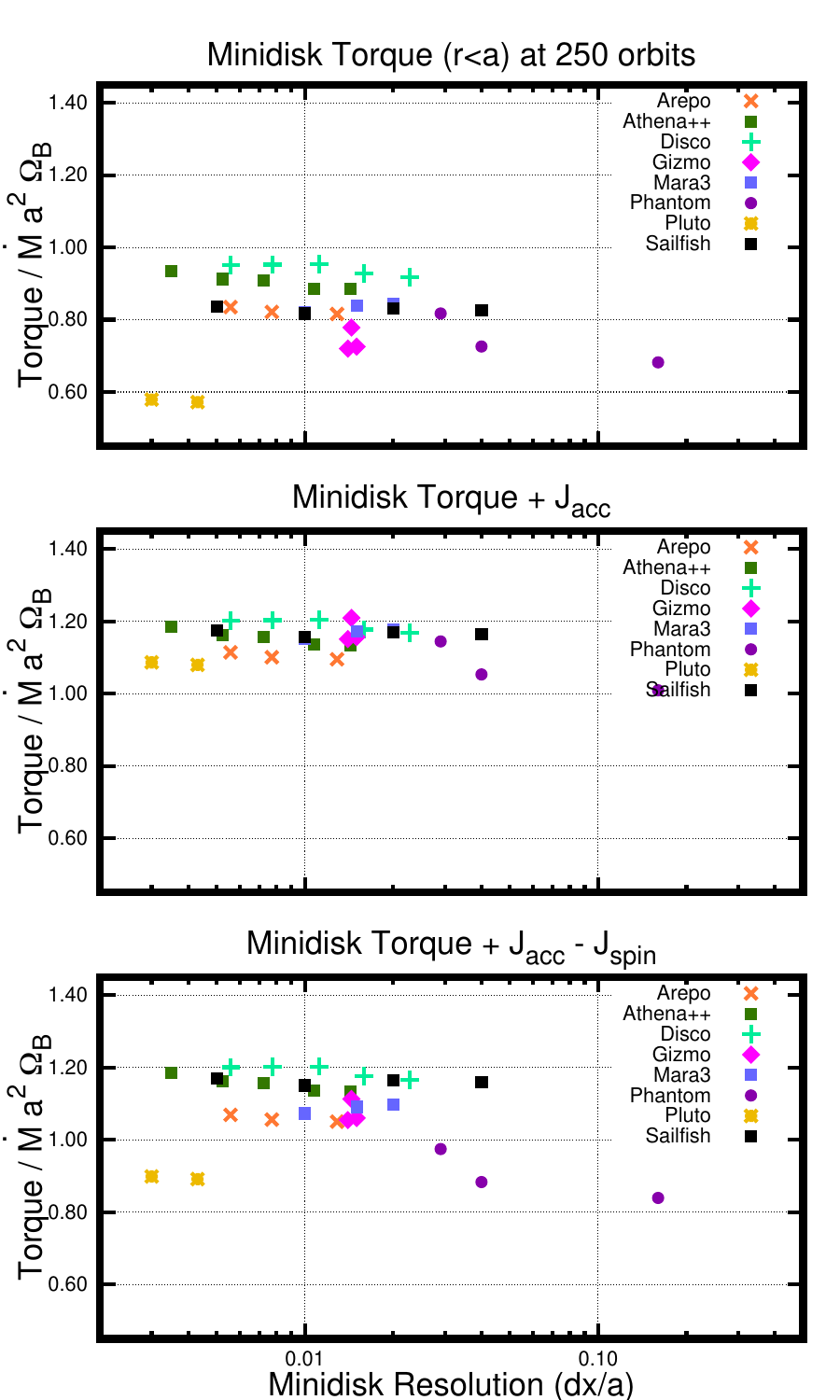}
\caption{Torques after being corrected by the sink term $\dot J_{\rm acc}$.  While different codes disagree on the value of the minidisk torque ($T_{\rm grav}$ within $r<a$; top panel) and they disagree on the angular momentum eaten by their sink prescription ($\dot J_{\rm acc}$), we find the sum $T_{\rm grav}+\dot J_{\rm acc}$ to converge to the same value for all codes (shown in the center panel).  Note, if we only include the "orbital angular momentum" accreted ($\dot J_{\rm acc} - \dot J_{\rm spin}$; bottom panel), the discrepancy persists.  Thus, to compute a orbital evolution of the binary that is consistent between all codes, one must simply take all angular momentum accreted by the sink (whether spin or orbital angular momentum) and give it to the orbital angular momentum of the binary.} 
\label{fig:jdot_corr}
\end{figure}

\begin{figure*}
\includegraphics[width=7in]{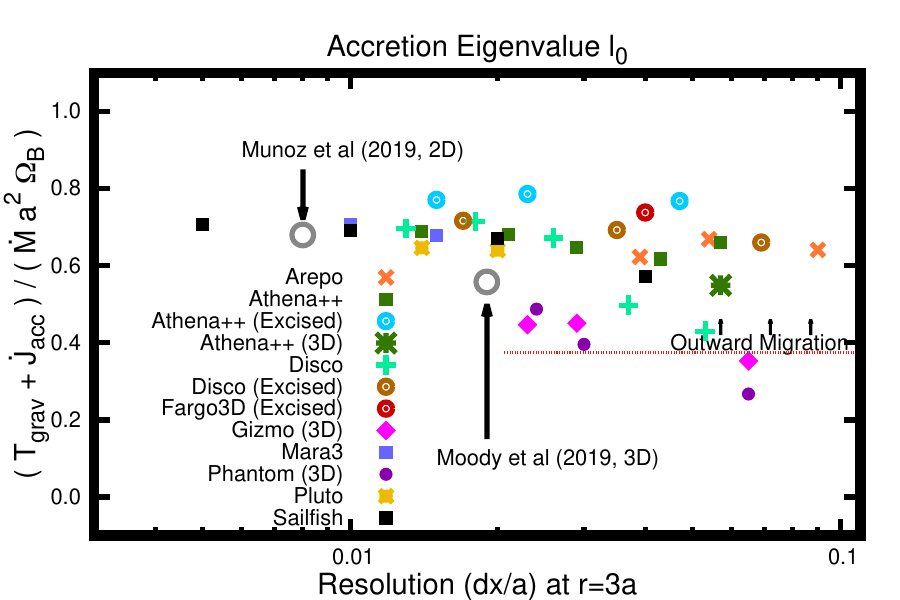}
\caption{Final comparison between all codes on the total angular momentum delivered to the binary, after including the sink term $\dot J_{\rm acc}$.  Vastly better agreement is found between codes after accounting for this contribution.  2D codes consistently converge toward an eigenvalue of $l_0 \approx 0.7$.} 
\label{fig:jtot_corr}
\end{figure*}

After accounting for the differences between 2D and 3D versions of the same problem, a (smaller but finite) discrepancy remains between different codes (see the lower panel of Figure \ref{fig:300torq}).  For codes running the same number of dimensions, the biggest difference in our methods is the sink prescription, which varies significantly between codes. It would be a major practical challenge to implement the same sink prescription across all codes; particle-based codes tend to remove gas instantaneously after a particle satisfies some conditions, whereas grid-based codes often employ a sink term to the equations, removing gas on a finite timescale.  Ideally our solutions would not depend on these choices at all, so the fact that different codes use different sinks is a concern that we address and quantify here.

After some experimentation, we found that our measure for the torque ($T_{\rm grav}$, Eq.~\ref{eqn:torque}) i.e. the moment arm crossed with gravitational force integrated over all fluid elements, is sink-dependent \citep[if sink particles are not treated carefully,][]{dittmann_sinks}.  The sink-dependence is because the total torque on the binary is partitioned between gravitational and accretion torques. When angular momentum is removed from the system by the sink, that angular momentum is no longer available to be exchanged with the binary via gravitational torques, and therefore it reduces the measured gravitational torque.

In the limit of very small sinks, the spin angular momentum removed $\dot J_{\rm spin}$ should be zero, and the orbital angular momentum removed should be $\dot J_{\rm orb} = \sum_i,\dot{M}_ij_i,$ or $0.25 \dot M a^2 \Omega_B$ for the equal-mass binaries studied here.  However, for finite-sized sinks (as all codes require) this can only be the case when using a careful sink treatment \citep[e.g.,][]{2020ApJ...892L..29D,dittmann_sinks}.  

We found that the different codes in our study with different sink prescriptions removed different amounts of both orbital and spin angular momentum from the system, leading to different predictions for the gravitational torque. It is also true in physical systems such as black hole accretion there will be a real $\dot J_{\rm spin}$ due to spin angular momentum removed at the ISCO, but for the purposes of this study we consider the ISCO to be at a much smaller radius than we are able to resolve.  However, this shows that such effects in real physical systems may be capable of altering the sign of the torque (by stealing angular momentum from the system and giving it to the spin of the binary instead of the orbit).

That sinks could artificially change the measured gravitational torque (sufficiently to change its sign) was first noted by \cite{tang17}. On the other hand, \cite{munoz19} computed the total angular momentum given to the binary by measuring the flow rate of angular momentum through the disk in steady-state, and argued that this measurement should be independent of the details of the sink prescription.

Many of the codes in our study are not set up to compute the angular momentum flux through the disk.  Nevertheless, we can compute the total angular momentum given to the binary by adding the gravitational torque to the accreted angular momentum per unit time (both spin and orbital contributions).  The hope is that while $\dot J_{\rm acc}$ and $T_{\rm grav}$ are sink-dependent, their sum might not depend on the sink.

In order to compute these diagnostics, we must define $\dot J_{\rm acc}$ in terms of the sink term $\dot \Sigma_{\rm sink}(\vec r,t)$.  Assume the hydrodynamics equations can be expressed as, %follows,
\begin{align}
    \partial_t \Sigma  + \nabla_i \left( \Sigma v_i \right) &= -\dot\Sigma_{\rm sink} \\
    \partial_t \left( \Sigma v_j \right) + \nabla_i \left( \Sigma v_i v_j + P \delta_{i j} \right) &= -\Sigma \nabla_i \Phi_{\rm grav} -  \dot\Sigma_{\rm sink} \vec{v}_*.
\end{align}
Then the sink term $\dot \Sigma_{\rm sink}$ contains information about the rate of removal of mass and angular momentum from the system,
\begin{equation}
    \dot M = \int \dot \Sigma_{\rm sink} dA,
\end{equation}

\begin{equation}
    \dot J_{\rm acc} = \int \dot \Sigma_{\rm sink} \vec r \times \vec v_* dA,
\end{equation}

where $\vec v_*$ is the velocity associated with the removed mass and momentum given above.  For "standard" sinks, gas is removed at the velocity of the fluid element, so $\vec v_* = \vec v_{\rm hydro}$.  However, for alternative sink methods, such as the "torque-free" sink in \texttt{Athena++} and in \texttt{Disco} \citep{2020ApJ...892L..29D, dittmann_sinks}, the velocity $v_*$ is chosen to avoid introducing spurious torques to the system. For \texttt{Athena++}, this is accomplished by boosting to the frame of the sink, and subtracting off the spin component of $\vec v_*$, i.e. only keeping a radial component in the frame of the sink.  For \texttt{Disco} (in this study), this is accomplished in a simpler manner, by setting $\vec v_* = \vec v_{\rm sink}$, so that $\vec v_* = 0$ in the frame of the sink (although this also affects the radial component of the velocity near each sink).  For the standard sinks employed in most codes, $\vec v_*$ is the local fluid velocity.  Note that particle-based codes accrete gas particles discretely, and in this study $\vec v_*$ is simply the velocity of the removed particle, i.e. the local fluid velocity.

Additionally, we have computed a separate quantity, the accreted spin,
\begin{equation}
    \dot{J}_{\rm spin} \equiv \int \dot{\Sigma}_{\rm sink} \left(\vec{r} - \vec{r}_s\right) \times \left(\vec{v}_* - \vec{v}_s\right) dA, \\
\end{equation}
where $\vec{r}_s$ and $\vec v_s$ are the position and the velocity of the sink particle.  One can then define the accreted {\em orbital} angular momentum:

\begin{equation}
    \dot{J}_{\rm orb} \equiv \dot{J}_{\rm acc} - \dot{J}_{\rm spin}
\end{equation}

Figure \ref{fig:jdot_corr} shows the minidisk torque in all our (non-excised) codes, after correcting for the sink term $\dot J_{\rm acc}$.  The top panel shows the raw minidisk torque (gravitational torque measured within $r<a$) in units of $\dot M a^2 \Omega_B$.  The center panel shows the minidisk torque plus the total angular momentum accreted, $T_{\rm mini} + \dot J_{\rm acc}$, again in units of $\dot M a^2 \Omega_B$, showing much better agreement between codes.  Note that this agreement disappears if one only includes $\dot J_{\rm orb} = \dot J_{\rm acc} - \dot J_{\rm spin}$.

How should we interpret this result?  Because each code handled accretion differently, the spin, accretion, and gravitational torques measured in each code depend on these numerical parameters (see Table \ref{tab:sinks}). However, the angular momentum current though the disk is largely set by viscosity and the gravitational influence of the binary on the disk. Then, thanks to Newton's third law and the conservation of angular momentum, this angular momentum current through the disk directly determines the total torque on the binary. Thus, while various codes find discrepant values of the gravitational, accretion, and spin torques, the measured values of the accretion eigenvalue $l_0$ tend to be in good agreement, regardless of accretion prescriptions or whether the binary was on the grid or excised.

Figure \ref{fig:jtot_corr} plots the accretion eigenvalue $l_0$ (Total angular momentum given to the binary in units of $\dot M a^2 \Omega_B$) for all codes at all resolutions in our study.  We find all 2D codes converge towards $l_0 \approx 0.7$, and that 3D codes converge towards a smaller value, $l_0 \approx 0.5$.  All of this is consistent with previous studies, which have found $l_0 = 0.68$ in 2D \citep{munoz19} and $l_0 = 0.56$ in 3D \citep{moody19}.  All eleven numerical techniques find outward migration for the binary in the relaxed state of the disk, so long as the code's resolution is sufficiently high.  In practice, this requires a resolution of around $\Delta x \lesssim 0.01a$ in the vicinity of the minidisks, and a resolution of $\Delta x \lesssim 0.03a$ at r = 3a for grid codes, while particle-based codes require higher resolution (by this metric) to achieve convergence.

\begin{deluxetable}{cccccc}
\tabletypesize{\footnotesize}
\tablewidth{0pt}  
\tablecaption{\label{tab:sinks} Sink Prescriptions for Non-Excised Codes} 
\tablehead{ \colhead{Code} & \colhead{$\gamma/\Omega_B$} & \colhead{ $R_S/a$} &  \colhead{Type} & \colhead{ $\dot J_{\rm acc}/
\dot M$} & \colhead{ $\dot J_{\rm spin}/
\dot M$} }
\startdata
\texttt{Arepo} & $\infty$\tablenotemark{a} & 0.03 & Standard & 0.2794 & 0.04489\\
\texttt{Athena++} & 1.33 & 0.05 & Torque-Free & 0.24996 & $10^{-20}$\\
\texttt{Disco} & 1 & 0.05 & Torque-Free & 0.2503 & 0.00075\\
\texttt{Gizmo} & $\infty$\tablenotemark{b}   & 0.05 & Standard & 0.4311 & 0.09650 \\
\texttt{Mara3} & 10 & 0.05 & Standard & 0.3332 & 0.08096 \\
\texttt{Phantom} & $\infty$\tablenotemark{b} & 0.05 & Standard & 0.3273 & 0.17258 \\
\texttt{PLUTO} & $10^3$ & 0.075 & Standard & 0.5073 & 0.18770\\
\texttt{Sailfish} & 10 & 0.05 & Standard & 0.3395 & 0.00602\\
\hline
\enddata

\tablecomments{\texttt{Fargo3D} was not included in this table, as this method excised the binary for all runs. \tablenotemark{a}{\texttt{Arepo} removed a fraction (up to $50\%$ nearest the sink particle) of the mass within each cell within the sink region every time step, resulting in a comparatively large but time-dependent sink rate.} \tablenotemark{b}{\texttt{Gizmo} and \texttt{Phantom} removed particles within the sink region instantaneously.}}
\end{deluxetable}

\subsection{Finite vs Infinite Disk}
\label{sec:finite}

\begin{figure}
\includegraphics[width=3.2in]{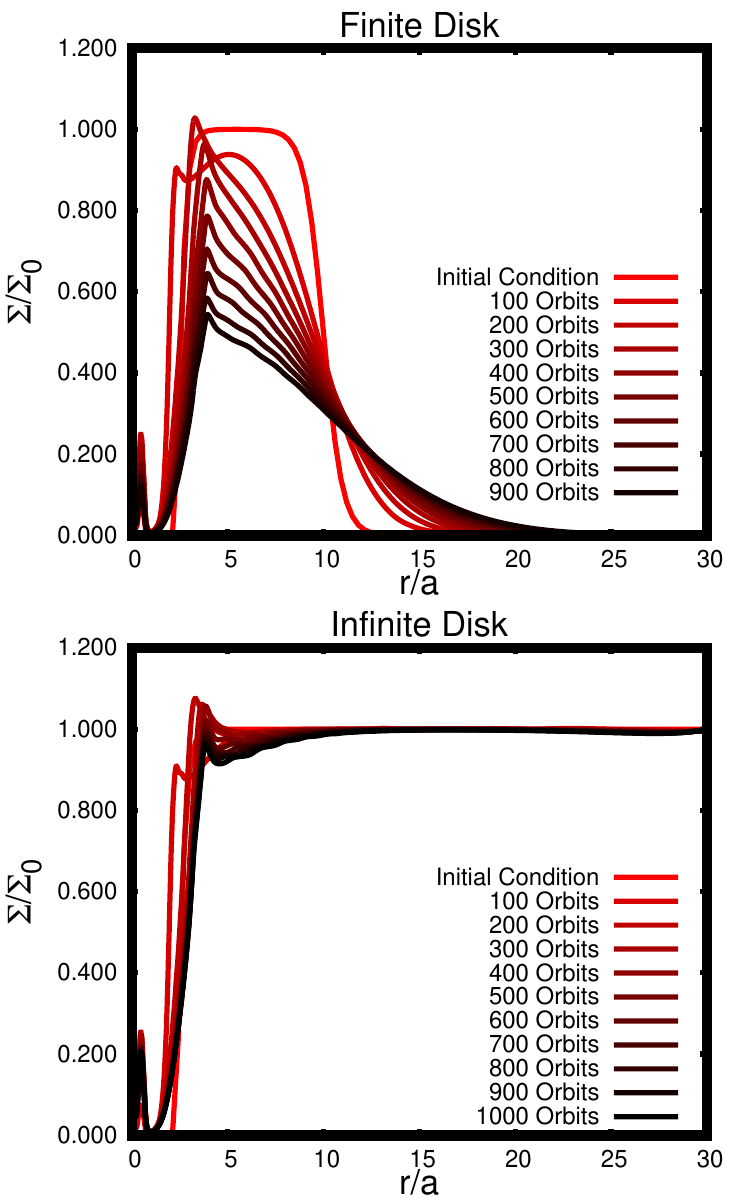}
\caption{Evolution of our two disk models over time in the \texttt{Disco} code.  During the first 100 orbits or so, the finite and infinite disks exhibit similar surface density.  However, as time evolves, the finite disk accretes mass and spreads, reducing the surface density to half its original value over 1000 orbits.  The infinite disk reaches a reasonably steady configuration at $\Sigma \sim \Sigma_0$.  At late times, the disk structure is completely different, but the torque (when normalized to $\dot M$) is nearly identical, as shown in Figure \ref{fig:finite_2}.} 
\label{fig:finite_1}
\end{figure}

\begin{figure}
\includegraphics[width=3.2in]{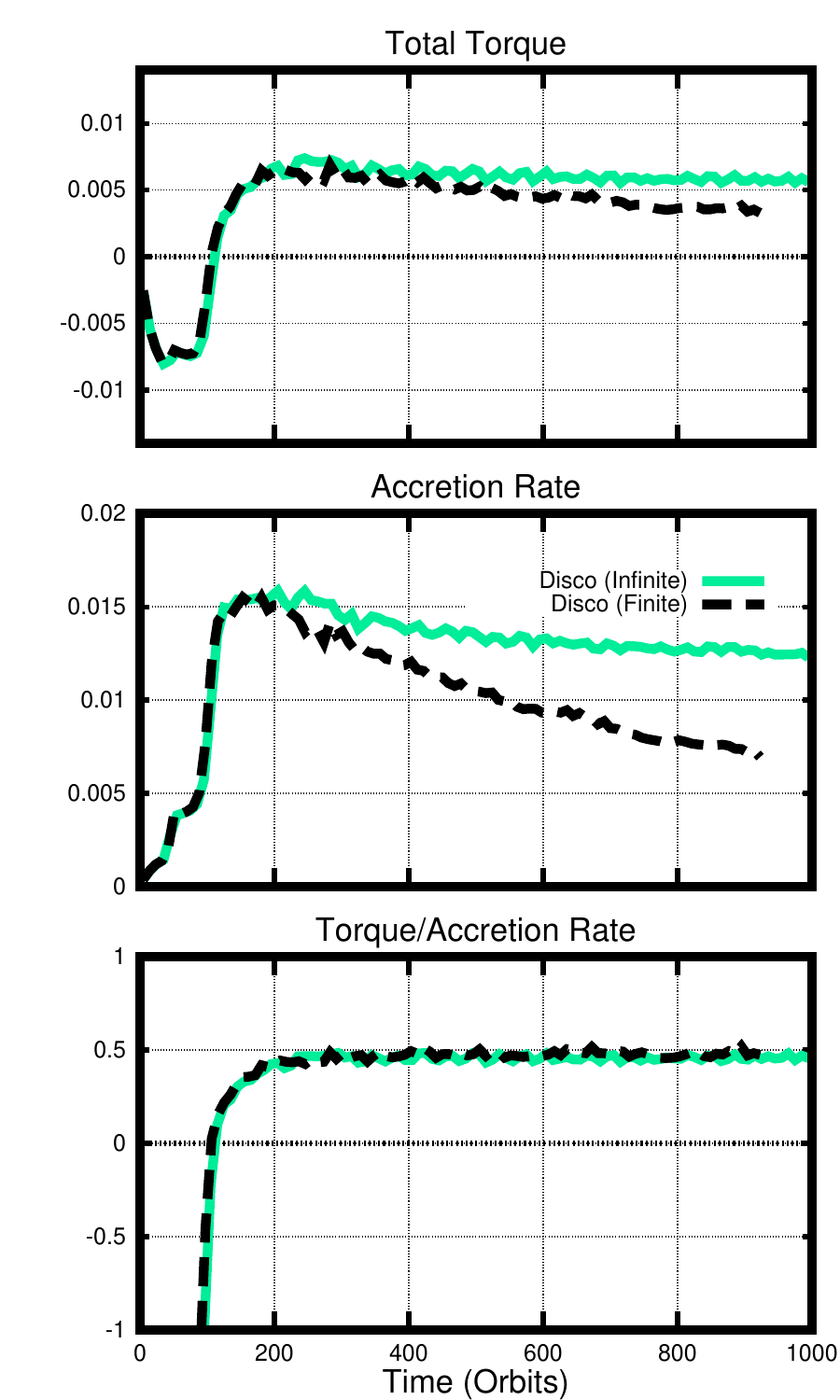}
\caption{Finite and Infinite disk models compared in the \texttt{Disco} code over 1000 orbits.  As shown in Figure \ref{fig:finite_1}, the disk structure at 1000 orbits is completely different, such that the torque and accretion rate are about half as large in the finite disk as they are in the infinite disk.  However, their ratio $T/\dot M$ is identical for the finite and infinite disk; this ratio does not appear to depend on disk structure at all, though a more systematic investigation of different disk models might be able to demonstrate this explicitly.} 
\label{fig:finite_2}
\end{figure}

The effect of a finite disk has already been investigated by \cite{munoz_finite}.  Nevertheless, we can empirically demonstrate any difference between a finite and infinite disk, by comparing the two choices for $f(r)$ in Equation \ref{eqn:fr}.  Surface density evolution of the two different models is plotted in Figure \ref{fig:finite_1}.  Torque as a function of time for the two models is plotted in Figure \ref{fig:finite_2}.  When one waits for a sufficiently long time, the torque and accretion rates are reduced in the finite-disk case, owing to the fact that the disk mass is being depleted and the disk is spreading.  However, even in this case, if one plots the ratio of torque to accretion rate, one achieves an identical answer between finite and infinite disks.  Therefore, both models provide an equivalent measurement of the torque.  

\subsection{Live vs Fixed Binary}

\begin{figure}
\includegraphics[width=3.3in]{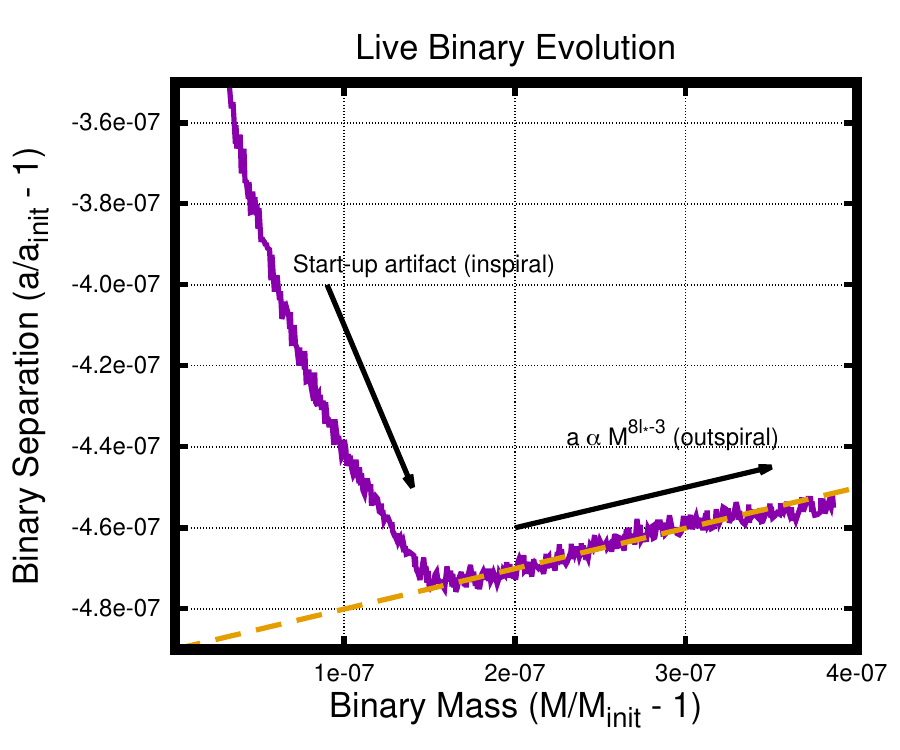}
\caption{Employing a "Live Binary" in the \texttt{Phantom} code. For sufficiently low surface density $\Sigma_0$, the impact of the live binary is negligible and the binary migrates according to Equation (\ref{eqn:a(M)}) (orange dashed line), with $l_* \approx 0.39$ consistent with the measured torques.  Additionally, there is an artificial start-up transient consistent with the negative torques seen in the first 150 orbits, highlighting the need for long-duration computations.} 
\label{fig:live}
\end{figure}

Most of these calculations employ a binary on a fixed circular orbit, assuming the limiting case where the binary orbital evolution is slow compared to the relaxation time of the disk.  Such a choice is only valid in the limit of a low disk mass, akin to a test particle in orbital dynamics. If one wishes to follow the interaction of a binary with a substantially massive disk, the binary orbit should react accordingly.
We measure whether this effect is significant for low disk mass by analyzing orbital evolution in the \texttt{Phantom} code run, where the orbit is not fixed a priori, but the disk mass is assumed to be negligible, i.e. $M_{\rm d}/M = 5\times 10^{-6}$, in order to minimize the change in binary orbit due to the interaction with the disk. 

For a circular equal-mass binary, the orbital angular momentum is
\begin{equation}
    J = 0.25 M a^2 \Omega_B = 0.25 M \sqrt{ G M a }.
\end{equation}
As the orbit and mass of the binary change, one can compute $\dot J = l_0 \dot M a^2 \Omega_B$ in terms of $\dot a$ and $\dot M$,
\begin{equation}
    \frac{\dot J}{J} = \frac32 \frac{\dot M}{M} + \frac12 \frac{\dot a}{a} = 4 l_0 \frac{\dot M}{M}.
\end{equation}
This reduces to an expression for $\dot a$ in terms of $\dot M$,
\begin{equation}
    \frac{\dot a}{a} = ( 8 l_0 - 3 ) \frac{\dot M}{M},
\end{equation}
which can be solved to determine the separation as a function of mass $a(M)$,
\begin{equation}
    a \propto M^{8 l_0 - 3}.
\end{equation}

In general, we expect our assumption to neglect variations of the binary orbit not to be valid when the disk mass is significant in comparison to the binary mass \citep[see, e.g.,][for a discussion]{Franchini2023, TiedeDZD:2024}.  In particular, at the resolution tested, with an eigenvalue $l_0=0.39$, we obtain the scaling $a(M) \propto M^{0.1}$, consistent with the numerical result by \texttt{Phantom}, as shown in Figure~\ref{fig:live}.
 
Note that an artificial period of inspiral occurs (consistent with the negative torque seen in all codes for the first 100 orbits).  This highlights the need for long-duration computations to ensure such artificial start-up transients do not dominate the solution.  This is important because in the first 100 orbits the solution should depend on our choice of initial density profile, whereas the late-time behavior does not. For example, Figures \ref{fig:finite_1} and \ref{fig:finite_2} show that very different disk models give nearly identical accretion eigenvalues $l_0$ in steady-state; this was also shown previously by \cite{munoz_finite}.

It should also be noted that the live binary setup in \texttt{Phantom} only includes the orbital angular momentum accreted $\dot J_{\rm orb}$ not the total accreted angular momentum -- i.e. it neglects the $\dot J_{\rm spin}$, assuming this term does not contribute to the total angular momentum of the binary (e.g. assuming the point mass is a black hole and the spin is accreted and spins up the hole).  So, for the live binary setup of \texttt{Phantom}, the expansion of the binary will follow

\begin{equation}
    a \propto M^{8 l_* - 3}.
    \label{eqn:a(M)}
\end{equation}
where
\begin{equation}
    l_* = \frac{T_{\rm grav} + \dot J_{orb}}{ \dot M a^2 \Omega_B}
\end{equation}
which explains the somewhat slower expansion rate than one would expect by including both orbital and spin angular momentum accreted. Nevertheless, the migration of the binary orbit does not affect the measured torques or accretion rates, and the migration is consistent with the gravitational torque combined with the accreted orbital angular momentum term. 

\begin{figure}
\includegraphics[width=3.3in]{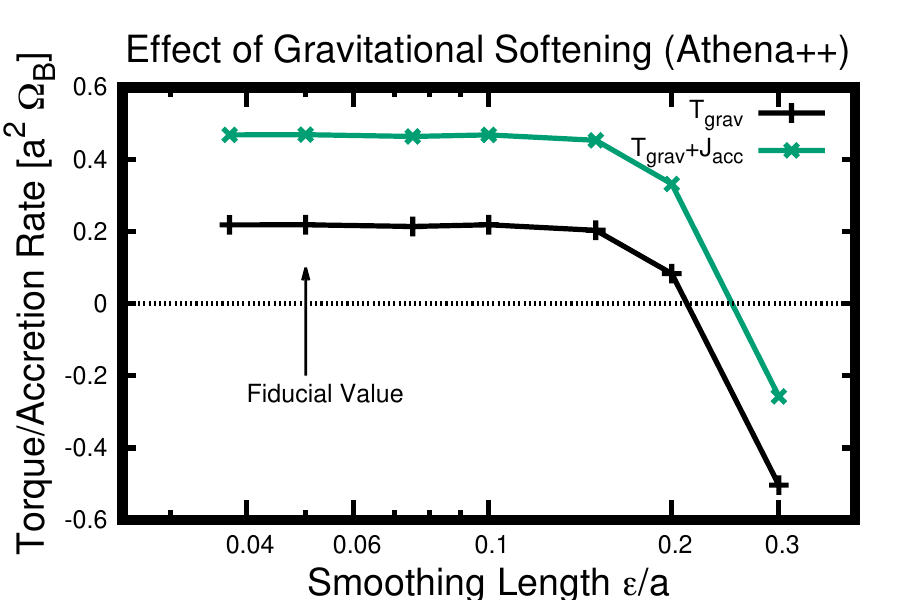}
\caption{Gravitational softening was varied in the \texttt{Athena++} code.  The value of $\epsilon$ represents the length scale being smoothed over.  This demonstrates that our fiducial value $\epsilon = 0.05a$ is much smaller than necessary to give converged torques.  The criterion turns out to be $\epsilon \lesssim 0.15 a$.  This criterion would likely vary with the Mach number, mass ratio, or eccentricity of the binary.} 
\label{fig:eps}
\end{figure}

\subsection{Gravitational Softening}
\label{sec:eps}
One approximation that was made for the sake of uniformity across codes was the form of the gravitational potential,
\begin{equation}
    \phi_i = \frac{G M_i}{\sqrt{ (\vec r - \vec r_i)^2 + \epsilon^2 }},
\end{equation}
with $\epsilon = 0.05$.  As a simple test of how much this affects our solutions, we ran this same problem using the \texttt{Athena++} code using a range of values of $\epsilon$.  The results of this series of calculations appear in Figure \ref{fig:eps}.  We find that for sufficiently small $\epsilon \lesssim 0.15a$, our measured torque is independent of the choice of $\epsilon$, suggesting that the gravitational softening does not affect our results.  This also suggests that the precise form of the potential did not matter in the vicinity of the point masses (as some studies favor cuspier potential forms e.g. using spline softening). A softening of $\epsilon \sim 0.15a$ is quite large, comparable to the size of the minidisks.  We did not test the eccentric case, but it is likely that this nice "plateau" behavior would occur at smaller scales if the binary were unequal mass or on an eccentric orbit, or orbiting retrograde to the disk \citep[see softening studies in][]{TiedeDOrazio:2023,dittmann_q}, so we can only say what value and form of the potential is sufficient for this specific test problem.  For example, the value of $\epsilon$ found in this study is sufficiently small that softening is negligible beyond the Roche lobe of each sink; naturally, this scale would become smaller near the pericenter of an eccentric orbit or around lower-mass objects. Regardless, for this test problem that our softening length is sufficiently small and, as shown in \citet{dittmann_sinks}, the precise form of the potential is immaterial.

\section{Conclusions}

We have developed a straightforward code test that exhibits the basic fundamental features of the binary-disk problem.  We have tested eleven very different numerical schemes on this same problem, to assess code performance and to test whether very different codes can agree on the solution to the same underlying physics problem.  Our code outputs have been made public online at \url{physics.purdue.edu/duffell/data.html} and on a permanent Zenodo repository \citep{sbb_repository}.  Although there are some expected differences between 2D and 3D setups, overall we find general agreement between codes, with small deviations that we believe can be accounted for or understood.

Over the first 1-10 orbits, all codes can achieve high-precision convergence (this includes both excised runs and calculations performed in 3D) and this work can therefore provide a convenient code test for checking ones initial setup, diagnostics, and code convergence.

Over the first 100 orbits, the disk becomes eccentric and agreement between codes becomes more difficult to determine.  We find agreement in the growth rate of the instability $\Gamma_c/\Omega_B \approx 0.012$ and the precession rate of the cavity $\Omega_c/\Omega_B \approx 0.004$.  Here is where we first see differences between 2D and 3D, as the cavity is smaller in 3D and the precession rate is faster (more like 0.045 / binary period).  The instability growth rate, however, seems to be comparable between 2D and 3D.

For the problem studied here, torque saturates to steady-state in a few hundred orbits. At this time, $\dot J / \dot M$ is independent of the disk size \citep[as also shown previously by][]{munoz_finite}, as demonstrated by Figures \ref{fig:finite_1} and \ref{fig:finite_2}. This suggests that this code test can be extended to more complicated and disk models. The final average gravitational torque was $T_{\rm grav} \approx 0.5 \dot M a^2 \Omega_B$, albeit with significant discrepancies between codes. 

The biggest contribution to this discrepancy was the fact that some codes were run in 2D, while others simulated these models in 3D.  By running \texttt{Athena++} in 3D and \texttt{Gizmo} in 2D, we were able to determine how much of the discrepancy was due to the dimensionality of the problem. This accounts for a discrepancy of $\delta T \approx 0.16 \dot M a^2 \Omega_B$, with 3D calculations experiencing a lower (but still positive) torque. This difference between 2D and 3D torque is consistent with the difference found by \citet{moody19}.

After accounting for this discrepancy, smaller discrepancies remained between codes, which were caused by the different sink prescriptions employed. However, the total angular momentum current through the disk should be largely independent of most sink choices, and thus the total torque on the binary should not depend on the sink prescription. This also leads to good agreement between excised codes and those which simulate the binary itself. This also shows that in a real physical system, if the point masses remove angular momentum from the disk and contribute it to the spin (e.g. if the black hole ISCO is sufficiently large) this could remove angular momentum from the circumbinary disk without giving it to the binary's orbit, affecting the rate or direction of an objects migration. 

After accounting for both dimension and sink discrepancies, there are still (much smaller) deviations in the torque between codes, which could reflect how precisely these codes are able to converge.  Nevertheless, all codes are sufficient to predict the direction of migration.

For this test problem, all codes found outward migration.  Most codes fixed the binary orbit and mass, but some used a live binary, which obeyed the scaling

\begin{equation}
    a \propto M^{8 l_0 - 3}
\end{equation}

consistent with the angular momentum eigenvalue 

\begin{equation}
    l_0 = \frac{T_{\rm grav} + \dot J_{\rm acc}}{\dot M a^2 \Omega_B}
\end{equation}

Thus, for sufficiently low disk mass, including the live reaction of the binary orbit to the gas forces is not vital for determining the orbital evolution of the system.  This can be understood in the limit where the disk is modeled as a "test fluid" and hence the orbital evolution of the binary is sufficiently slow that it can be well-approximated by a fixed orbit. 

Overall, we found 2D codes converged to an eigenvalue of $l_0 \approx 0.7$ (with 3D codes converging to a lower value, $l_0 \approx 0.5$), and that this eigenvalue was very consistent across all 2D codes, so long as accreted angular momentum was accounted for.  Codes that excised the binary were able to recover this eigenvalue so long as the angular momentum flow through the excision radius was accounted for.

\acknowledgements
This research was supported in part by grant NSF PHY-1748958 to the Kavli Institute for Theoretical Physics (KITP).  Foundational meetings initiating this research took place at the KITP, during the program "Bridging the Gap: Accretion and Orbital Evolution in Stellar and Black Hole Binaries" (https://www.kitp.ucsb.edu/activities/binary22).

%%personal ackn in author order:
PD is supported by the National Science Foundation under grant No. AAG-2206299.
AJD acknowledges the University of Maryland supercomputing resources (http://hpcc.umd.edu) that were made available for conducting the research reported in this paper, and the ASTRA computing cluster maintained by the Department of Astronomy at the University of Maryland. 
A.J.D. was supported in part by NASA ADAP grant 80NSSC21K0649.
D.J.D. received funding from the European Union's Horizon 2020 research and innovation programme under Marie Sklodowska-Curie grant agreement No. 101029157, and from the Danish Independent Research Fund through Sapere Aude Starting Grant No. 121587. The Tycho supercomputer hosted at the SCIENCE HPC center at the University of Copenhagen was used for supporting this work. 
AF acknowledges financial support provided under the European Union’s H2020 ERC Consolidator Grant "Binary Massive Black Hole Astrophysics" (B Massive, Grant Agreement: 818691).
E.R. thanks Daniel Price, Guillaume Laibe and Elliot Lynch for fruitful discussions and acknowledges financial support from the European Union's Horizon Europe research and innovation programme under the Marie Sk\l{}odowska-Curie grant agreement No. 101102964 (ORBIT-D), and from the European Research Council (ERC) under the European Union’s Horizon 2020 research and innovation programme (grant agreement No 864965, PODCAST).  J. Zrake acknowledges support from the \emph{LISA} Preperatory Science Program (LPS) through NASA Award No. 80-NSSC-24K0440.

Numerical calculations were performed using many different computing facilities, including the \texttt{Petunia} cluster at Purdue University. \texttt{Sailfish} simulations were run on the Palmetto cluster at Clemson University. \texttt{Phantom} simulations were performed using the DiRAC Data Intensive service at Leicester, operated by the University of Leicester IT Services, which forms part of the STFC DiRAC HPC Facility (www.dirac.ac.uk). The equipment was funded by BEIS capital funding via STFC capital grants ST/K000373/1 and ST/R002363/1 and STFC DiRAC Operations grant ST/R001014/1. DiRAC is part of the National e-Infrastructure.

Research presented in this article was supported by the Laboratory Directed Research and Development program of Los Alamos National Laboratory under project No. 20220087DR. This research used
resources provided by the Los Alamos National Laboratory Institutional Computing Program, which is supported by the U.S. Department of Energy National Nuclear Security Administration under contract No. 89233218CNA000001. The LA-UR number is LA-UR-23-26505. 

Research at Perimeter Institute is supported in part by the Government of Canada through the Department of Innovation, Science and Economic Development and by the Province of Ontario through the Ministry of Colleges and Universities.

\texttt{PLUTO} simulations were performed with the support of the High Performance and Cloud Computing Group at the Zentrum f\"ur Datenverarbeitung of the University of T\"ubingen, the state of Baden-W\"urttemberg through bwHPC and the German Research Foundation (DFG) through grant no INST 37/935-1 FUGG. ABTP acknowledge support from the Royal Society in the form of a University Research Fellowship and Enhanced Expenses Award and grant 285676328 of the German Research Foundation (DFG).

Finally, we would like to thank the anonymous referee for the very thoughtful and helpful report.

\bibliography{refs}

\appendix
\section{Code Descriptions}\label{sec:codeAppendix}
In the following, we provide details for each code, and the various algorithmic choices employed in this work. 
\subsection{Arepo}

\texttt{Arepo} \citep{arepo} is a moving-mesh hydrodynamics code based on a dynamic Voronoi tesselation of the computational domain performed each timestep.  \texttt{Arepo} was originally designed for cosmological simulations but has been adapted for the solution of the Navier-Stokes equations \citep{2013MNRAS.428..254M} and the study of astrophysical disks \citep[e.g.,][]{MunozLai:2016}, especially protoplanetary disks and circumbinary disks. The mesh motion helps by performing hydrodynamic calculations in the frame of the otherwise supersonic (Mach 10 in this study) flow. Additionally, \texttt{Arepo}'s resolution is naturally adaptive in nature, resulting in much higher resolution in the minidisks surrounding each point mass than further out in the disk (the zone size near each binary component is nearly an order of magnitude smaller than it is at a radius of a few times the binary separation).

\subsection{Athena++}
\texttt{Athena++} is a finite-volume code, supporting Cartesian, cylindrical, and spherical geometries in addition to both static and adaptive mesh refinement \citep{2020ApJS..249....4S}. \texttt{Athena++} is a rewrite of the \texttt{Athena} code \citep{2008ApJS..178..137S} using C++, which introduced new features such as block-based adaptive mesh refinement \citep[e.g.,][]{10.1145/509593.509650} and dynamic task scheduling. In this work, \texttt{Athena++} has been used to solve the equations of viscous hydrodynamics in both two and three dimensions, using both Cartesian coordinates with the binary on the grid, and cylindrical coordinates with an excised binary.

The 2D Cartesian simulations used the Roe approximate Riemann solver \citep{1981JCoPh..43..357R}, third-order piecewise parabolic spatial reconstruction \citep{2018JCoPh.375.1365F}, and the second-order van Leer (VL2) predictor-corrector time integrator described in \citep{2009NewA...14..139S}. We also carried out a limited number of three-dimensional simulations in Cartesian geometry, utilizing piecewise linear spatial reconstruction \citep{1974JCoPh..14..361V}, the VL2 time integrator, and HLLC approximate Riemann solver \citep{1994ShWav...4...25T}.

\subsection{Disco}

\texttt{Disco} is a finite-volume code that employs a moving, shearing polar grid \citep{2016ApJS..226....2D}.  The numerical method is essentially the "moving mesh" technique of \texttt{Arepo}, but rather than using Voronoi cells, \texttt{Disco} employs moving volumes that are wedge-like annular segments (as in a fixed polar grid code) which can shear azimuthally.  The moving mesh reduces advection errors as it does in \texttt{Arepo}, but the smooth shearing of the zones reduces numerical noise, which can be advantageous for many problems, especially where subtle or weak nonlinear effects must be evolved and measured alongside the large bulk Keplerian shear flow of the disk.

In this comparison, \texttt{Disco} was run in two configurations, one in which the grid extended to $r=0$, and another with an excised inner region. In the former, the grid spacing was logarithmic at large radii and linear near the origin, and the azimuthal resolution was varied in each annulus to maintain a roughly constant aspect ratio. In the latter configuration, a diode boundary condition was applied at the excision radius ($r=a$, where $a$ is the binary separation). Notably, both versions of \texttt{Disco} used in this study employed a simplified version of the viscous terms in the hydro equations \citep[see][Section 2.9 and the appendix, for the precise equations being integrated]{2016ApJS..226....2D}.  When compared with a version of the code that includes all viscosity terms in the equations, one finds only minor deviations for equal-mass binaries such as those studied in this comparison \citep[see Appendix A of][]{dittmann_sinks}, but larger deviations have been found for unequal-mass binaries \citep[see Appendix B of][]{dittmann_q}.

\subsection{Fargo3D}

\texttt{Fargo3D} \citep{FARGO3D}, the successor of the \texttt{FARGO} code, is a versatile multifluid HD/MHD code that runs on clusters of CPUs or GPUs, with special emphasis on protoplanetary disks.  An important aspect of \texttt{Fargo3D}, which makes it ideal for the study of protoplanetary disks, is the orbital advection scheme originally developed in the \texttt{FARGO} code \citep{fargo}, for use with tracking highly supersonic orbital motions. The code's name being \texttt{Fargo3D} may cause some confusion as it has only been run in 2D in this study (though as the name suggests it is capable of 3D). Distinguishing features of \texttt{FARGO3D} include its use of a staggered grid with face-centered velocity components and an artificial von Neumann-Richtmyer artifical viscosity as in \citep{1992ApJS...80..753S}.

\subsection{Gizmo}

\texttt{Gizmo} is a multi-methods and multi-physics code for hydrodynamics simulations which implements smoothed-particle hydrodynamics (SPH), meshless finite volume \citep[MFV,][]{2011MNRAS.414..129G}, and meshless finite mass schemes \citep{Hopkins2015}. The latter are very similar to moving-mesh methods, except that the Riemann problem is solved across smooth boundaries ("faces") between particles instead of using the faces resulting from a Voronoi tessellation. In particular, the method employed here is the Meshless Finite Mass (MFM) one, which implicitly assumes a deformation of the face between particles such that there is no mass flux between them, effectively making the method Lagrangian.

In this study, \texttt{Gizmo} has been used coupled with adaptive particle-splitting for numerical refinement of the gas dynamics inside the disk cavity \citep[see][for details]{Franchini2022,Franchini2023}. The kernel size for the hydrodynamic interaction is defined by an effective number of neighbours equal to 58 \citep{Franchini2022}.
The binary is modelled using two sink particles \citep{Bate1995} each with sink radius $r_{\rm sink}=0.05a$ (particles within this radius are removed from the simulation). 

\subsection{Mara3 and Sailfish}
\texttt{Mara3} and \texttt{Sailfish} both use a fixed-mesh 2nd-order Godunov solver. They solve the locally isothermal and vertically averaged Navier-Stokes equations in Cartesian coordinates. \texttt{Mara3} is an older code originally configured to study relativistic MHD turbulence \citep{Mara}. Features have been added specifically for simulations of the binary accretion problem, including the locally isothermal equation of state, viscous fluxes (constant-$\nu$ or constant-$\alpha$), fixed mesh refinement, test particles, and post-processing. \texttt{Mara3} was used to calculate gas-driven orbital evolution in \cite{Tiede:2020} and \cite{Zrake:2021}, and to study kinematic aspects of binary accretion using test particles in \cite{Tiede:2022}. \texttt{Sailfish} is a GPU-accelerated version of \texttt{Mara3} that also includes post-processing modules to synthesize light curves of thermal disk emission. It is the code that was used in \cite{WesternacherSchneider2022}, \cite{Krauth+2023}, and Westernacher-Schneider et al. (2023; in prep). \texttt{Sailfish} is very fast ($\gtrsim 10^9$ zone updates per second for this problem on an A100 GPU) and it runs efficiently on multi-GPU nodes.

\subsection{Phantom}

\texttt{Phantom} is a smoothed particle hydrodynamics (SPH, \citealp{lucy1977, gingold1977,monaghan1992}) and magnetohydrodynamics (SPMHD, \citealp{price2012}) three dimensional code \citep{Price2018}, developed for applications in the context of stellar, galactic, planetary and high energy astrophysics. The code has been used widely for studies of accretion disks (e.g. \citealp{lodato2010}) and turbulence (e.g. \citealp{pricefederrath2010}), from the birth of planets (e.g. \citealp{dipierro2015}) to how black holes accrete (e.g. \citealp{nealon2015,RagusaLodato:2016}), and more specifically for circumbinary disks (e.g. \citealp{nixon2013, facchini2013,ragusa2017, Franchini2019, Smallwood2019,hirsh2020,Franchini2021}).

In this work, the binary is modeled using two sink particles (\citealp{Bate1995,nixon2013} particles are removed from the simulation when they cross $r_{\rm sink}=0.05\,a$ and other criteria are met). Viscosity is calculated including the viscous stress term using two first derivatives \citep{Price2018} and constant $\nu=10^{-3}$, as prescribed for this code comparison setup. The artificial viscosity parameter for each particle is allowed to vary between $0.1 <\alpha_{\rm AV,i}<1$ using the \citet{Cullen2010} switch for shock capturing and $\beta_{\rm AV}=2$ to avoid particle interpenetration. Individual particle timestepping has been used to speed up the simulations. For this work, 5 simulations differing for their resolution have been performed (using 1M, 10M, 20M, 50M, 100M particles). Among them, only those with 1M, 10M and 20M particles reached 300 binary orbits. Results from the 50M and 100M particles runs will be used for diagnostics in the first 10-15 binary orbits. The resolution $\Delta x/a$ used to compare the various diagnostics have been calculated as the mode of the distribution of the particle smoothing lengths in the region of reference prescribed. 

%\cc{AJD: Please forgive my ignorance, but there was quite a discussion on SPH artificial viscosity in \citet{lodato2010}... are the choices here consistent with what they found necessary for reproducing analytical theory?}
%\cc{ER: Sure. To clarify, I am not using artificial viscosity set to be a form of  ``physical'' viscosity for transporting angular momentum, as discussed in \citet{lodato2010}. Artificial viscosity is used only for shock capturing, which implies keeping it as low as possible except in shock regions; this can be achieved with a viscosity switch \citep{Cullen2010}, which increases the value of $\alpha_{AV}$ only for particles close to shocks. The physical viscosity is included solving Navier-Stokes equation. In this specific instance the stress term is calculated using two first derivatives of the velocity field (instead of performing one second derivative, but this is just numerical choice about how to implement the viscous stress tensor). With our choice of $\nu$, which is relatively large, and large number of particles the $\alpha_{AV}$ contribution to physical viscosity is negligible compared to that coming from $\nu$. I had forgot to include a sentence about physical viscosity. I added it in the text.}

\subsection{PLUTO}

\texttt{PLUTO} is a finite-volume grid code developed by \cite{2007Mignone} using Cartesian, cylindrical, or spherical grids. \texttt{PLUTO} was created to solve astrophysical problems that require hydrodynamical, magneto-hydrodynamical and relativistic-hydrodynamical prescriptions. The version used in this project is a GPU-capable modification by \cite{2017Thun}. The numerical setup generally follows \cite{2017Thun}. We use a HLL-sovler and a VL2 for the simulations. The binary is modelled in a 2D polar configuration with a radially logarithmic spaced grid up to $0.3~a$ and another 30 cells linearly spaced down to $0.03~a$. The inner boundary reflects radial velocities and density, such that no mass is lost at the inner radius near the origin.

\end{document}